\documentclass[%
 aip,
 amsmath,amssymb,
 reprint,%
]{revtex4-1}
\usepackage[utf8]{inputenc} %
\usepackage[T1]{fontenc}    %
\usepackage[colorlinks,
    linkcolor=citelightblue,citecolor=citelightblue,urlcolor=citelightblue]{hyperref}       %
\usepackage{booktabs}       %
\usepackage[table,xcdraw]{xcolor}         %

\usepackage{graphicx}

\usepackage{threeparttable}

\usepackage{multirow}

\usepackage{amsmath}
\usepackage{bm}
\usepackage{bbm}
\usepackage{amsthm}
\usepackage{amsfonts}
\usepackage{pifont}
\usepackage{enumitem} %
\usepackage[subrefformat=parens,labelformat=parens]{subfig}

\usepackage{wrapfig}
\usepackage[ruled, vlined, linesnumbered, noend]{algorithm2e}

\definecolor{citecolor}{RGB}{34,139,34}
\definecolor{mydarkblue}{rgb}{0,0.08,1}
\definecolor{mydarkgreen}{rgb}{0.02,0.6,0.02}
\definecolor{mydarkred}{rgb}{0.8,0.02,0.02}
\definecolor{mydarkorange}{rgb}{0.40,0.2,0.02}
\definecolor{mypurple}{RGB}{111,0,255}
\definecolor{myred}{rgb}{1.0,0.0,0.0}
\definecolor{mygold}{rgb}{0.75,0.6,0.12}
\definecolor{myblue}{rgb}{0,0.2,0.8}
\definecolor{mydarkgray}{rgb}{0.,0.2,0.2}

\definecolor{lightred}{RGB}{255,235,235}
\definecolor{lightgreen}{RGB}{235,255,235}
\definecolor{lightblue}{RGB}{235,235,255}
\definecolor{citelightblue}{RGB}{49,164,222}
\definecolor{lightcyan}{RGB}{235,255,255}
\definecolor{lightmagenta}{RGB}{255,235,255}
\definecolor{lightyellow}{RGB}{255,255,235}

\definecolor{qxkcolor}{RGB}{215,235,255}
\definecolor{softmaxcolor}{RGB}{230,235,255}
\definecolor{probxvcolor}{RGB}{255,255,235}

\definecolor{topkcolor}{RGB}{255,235,235}
\definecolor{zecolor}{RGB}{255,255,235}
\definecolor{dynacolor}{RGB}{235,255,255}

\definecolor{reviewcolor}{RGB}{0,0,200}

\newcommand{\calN}{\mathcal{N}}

\newcommand{\calQ}{\mathcal{Q}}

\theoremstyle{plain}

\theoremstyle{definition}

\newcommand{\ours}{\texttt{TeMPO}\xspace}
\newcommand{\oursd}{\texttt{TeMPO-D}\xspace}
\newcommand{\ourse}{\texttt{TeMPO-E}\xspace}

\usepackage{graphicx}%
\usepackage{dcolumn}%
\usepackage{bm}%

\usepackage[utf8]{inputenc}
\usepackage[T1]{fontenc}
\usepackage{mathptmx}
\usepackage{etoolbox}

\makeatletter
\def\@email#1#2{%
 \endgroup
 \patchcmd{\titleblock@produce}
  {\frontmatter@RRAPformat}
  {\frontmatter@RRAPformat{\produce@RRAP{*#1\href{mailto:#2}{#2}}}\frontmatter@RRAPformat}
  {}{}
}%
\makeatother
\begin{document}

\title[]{TeMPO: Efficient \underline{T}im\underline{e}-\underline{M}ultiplexed Dynamic \underline{P}hotonic Tensor C\underline{o}re for Edge AI with Compact Slow-Light Electro-Optic Modulator
}
\author{Meng Zhang}%
\thanks{Meng Zhang and Dennis Yin are equal contributors to this work and designated as co-first authors.}
\affiliation{ 
Department of Electrical, Computer, and Systems Engineering, Rensselaer Polytechnic Institute, Troy, NY 12180, USA%
}%
\author{Dennis Yin}%
\thanks{Meng Zhang and Dennis Yin are equal contributors to this work and designated as co-first authors.}
\affiliation{ 
School of Electrical, Computer and Energy Engineering, Arizona State University, Tempe, AZ 85287, USA%
}%
\author{Nicholas Gangi}%
\affiliation{ 
Department of Electrical, Computer, and Systems Engineering, Rensselaer Polytechnic Institute, Troy, NY 12180, USA%
}%
\author{Amir Begovi\'{c}}%
\affiliation{ 
Department of Electrical, Computer, and Systems Engineering, Rensselaer Polytechnic Institute, Troy, NY 12180, USA%
}%
\author{Alexander Chen}%
\affiliation{ 
Department of Electrical, Computer, and Systems Engineering, Rensselaer Polytechnic Institute, Troy, NY 12180, USA%
}%
\author{Zhaoran Rena Huang$^\ast$}%
\email{huangz3@rpi.edu}
\affiliation{ 
Department of Electrical, Computer, and Systems Engineering, Rensselaer Polytechnic Institute, Troy, NY 12180, USA%
}%
\author{Jiaqi Gu$^\ast$}%
\email{jiaqigu@asu.edu.}
\affiliation{ 
School of Electrical, Computer and Energy Engineering, Arizona State University, Tempe, AZ 85287, USA%
}%

\date{\today}%

\begin{abstract}
Electronic-photonic computing systems offer immense potential in energy-efficient artificial intelligence (AI) acceleration tasks due to the superior computing speed and efficiency of optics, especially for real-time, low-energy deep neural network (DNN) inference tasks on resource-restricted edge platforms.
However, current optical neural accelerators based on foundry-available devices and conventional system architecture still encounter a performance gap compared to highly customized electronic counterparts.
To bridge the performance gap due to lack of domain specialization, we present a time-multiplexed dynamic photonic tensor accelerator, dubbed \ours, with cross-layer device/circuit/architecture customization.
At the device level, we present foundry-compatible, customized photonic devices, including a slow-light electro-optic modulator with experimental demonstration, optical splitters, and phase shifters that significantly reduce the footprint and power in input encoding and dot-product calculation. 
At the circuit level, partial products are hierarchically accumulated via parallel photocurrent aggregation, lightweight capacitive temporal integration, and sequential digital summation, considerably relieving the analog-to-digital conversion bottleneck.
We also employ a multi-tile, multi-core architecture to maximize hardware sharing for higher efficiency.
Across diverse edge AI workloads, \ours delivers digital-comparable task accuracy with superior quantization/noise tolerance. 
We achieve a 368.6 TOPS peak performance, 22.3 TOPS/W energy efficiency, and 1.2 TOPS/mm\textsuperscript{2} compute density, pushing the Pareto frontier in edge AI hardware. 
This work signifies the power of cross-layer co-design and domain-specific customization, paving the way for future electronic-photonic accelerators with even greater performance and efficiency.
\end{abstract}

\maketitle

\section{Introduction}
\label{sec:Introduction}

\begin{figure}[]
    \centering
    \includegraphics[width=\columnwidth]{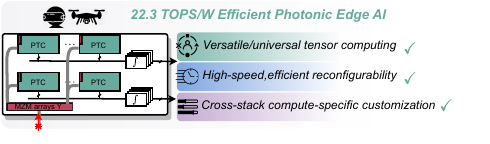}
    \vspace{-10pt}
    \caption{Our versatile, reconfigurable, cross-stack customized photonic accelerator \ours achieves digital-comparable accuracy with 22.3 TOPS/W efficiency on edge AI.}
    \vspace{-10pt}
    \label{fig:Teaser}
\end{figure}

Photonic computing has emerged as a promising technology for high-performance and energy-efficient computing, particularly in computation-intensive artificial intelligence (AI) tasks.
Various integrated photonic tensor core (PTC) designs have been introduced and demonstrated for ultra-fast photonic analog linear operation acceleration. 
Coherent PTCs that leverage interference and diffraction include MZI arrays~\cite{NP_NATURE2017_Shen}, butterfly-style meshes~\cite{NP_ASPDAC2020_Gu, NP_ACSPhotonics2022_Feng}, auto-designed photonic circuits~\cite{NP_DAC2022_Gu}, coupler-crossbar array~\cite{NP_arXiv2023_Zhu}, star-coupler-based design~\cite{NP_NatureComm2022_Zhu}, and metalens-based diffractive PTCs~\cite{NP_NatureComm2022_Wang}, etc.
Besides, to leverage the wavelength-division multiplexing (WDM) technique, there are incoherent multi-wavelength PTCs, e.g., MRR weight bank~\cite{NP_SciRep2017_Tait, NP_DATE2019_Liu, NP_DATE2020_Zokaee, NP_TCAD2020_Gu}, PCM crossbar arrays~\cite{NP_Nature2021_Xu}, micro-comb-based computing engine~\cite{NP_Nature2021_Feldmann,NP_NatureComm2023_Bai}.
We emphasize three key features of efficient PTCs required by general edge AI from the perspective of versatility, dynamic reprogrammability, and domain-specific customization, respectively, shown in Fig.~\ref{fig:Teaser}.

Versatility, or universality, is one of the important features of photonic AI hardware to accelerate a variety of DNN workloads.
A versatile/generic photonic accelerator based on universal optical linear units is capable of realizing general matrix multiplication (GEMM) and thus directly implementing a wide spectrum of pre-trained digital DNNs. 
Many specialized linear units are not applicable to generic tensor computation since they restrict their matrix expressivity to a subspace of specialized matrices for higher hardware efficiency, e.g., butterfly meshes~\cite{NP_ACSPhotonics2022_Feng} and tensorized MZI arrays~\cite{NP_APLPhotonics2021_Xiao}.

Besides versatility, photonic computing requires real-time, efficient input tensor encoding with low reconfiguration costs. 
One example is the MZI arrays, which support arbitrary weight matrices but suffer from high weight encoding costs due to the high complexity of matrix decomposition required to encode weights. 
Similarly, many subspace linear unit designs can approximate GEMM operations by cascading more programmable devices but require an even more costly optimization-based approach to map the weight matrix~\cite{NP_ACSPhotonics2022_Feng,NP_NatureComm2022_Zhu}.
Such a property restricts those designs to only support weight-static linear operations, e.g., fully-connected (FC) layers and convolutional (CONV) layers, where weights are pretrained and pre-encoded into the device/circuit transmissions.
However, advanced AI models, e.g., Transformer~\cite{NN_ICLR2021_Alexey, NN_ECCV2020_Carion, NN_ICML2021_Touvron,NN_NAACL2019_Kenton, NN_Neurips2020_Brown, openai2023gpt4} based on attention operations where both matrix multiplication operands are dynamic, full-range, and general tensors, cannot be efficiently mapped to those weight-static PTCs.

The third critical feature to enable efficient, scalable PTCs is domain-specific hardware customization.
\ding{202}~At the device level, many optical computing hardware demonstrations are based on standard foundry PDK elements, which are designed for optical communications and not optimized for analog neuromorphic computing.
For example, bulky electro-optic (E-O) modulators ($\sim$mm-level in length)~\cite{NP_OFC2018_Timurdogan} can be used as the transmitter module for high-speed communication but are not suitable for analog computing as the footprint is intractable with quadratically many such modulators for input encoding.
On the other hand, thermo-optic MZI modulators are usually compact but can only be modulated at KHz frequency due to the $\sim$10 $\mu s$ thermal constant and are usually power-consuming.
Plasmonic devices~\cite{NP_SciRep2021_Amin} are compact and high-speed but show high insertion loss ($>$10 dB), leading to significant laser power consumption.
Hence, compact, low-power, low-loss, and high-speed modulators are in high demand for efficient optical computing.
MRRs are compact and low-loss; however, their high locking power and high sensitivity to thermal variations limit their efficiency and robustness~\cite{NP_TCAD2021_Mirza}.
To bridge the gap at the device level, it is necessary to customize computing-specific optical components, e.g., multi-operand devices for compact neural computing~\cite{NP_TCAD2022_Gu, NP_DATE2021_Gu2}, diffractive meta-computing systems~\cite{ NP_NatureComm2022_Wang}.
\ding{203}~At the circuit level, customization is critical to reducing the long-lasting analog-to-digital and optical-to-electrical conversion bottlenecks.
\ding{204}~At the architecture level, due to the lack of optical memory, the large spatial footprint of photonic circuits, and the high digital memory access cost, the architecture topology and dataflow also need to be customized to fully leverage the temporal locality to reduce data movement cost and maximize hardware sharing.
Only with device-circuit-architecture cross-layer co-design and customization can we realize photonic computing's advantages compared to its electronic counterparts. 

In this work, we present a time-multiplexed dynamic photonic tensor accelerator design, dubbed \ours, for efficient edge AI acceleration, featuring ultra-compact slow-light electro-optic modulators for input operand encoding, hierarchical partial product accumulation with lightweight capacitive temporal integration modules and multi-core architecture to maximize sharing of data input/readout circuitry. 
One key innovation of this work is the utilization of custom-designed, foundry-fabricated slow-light MZI modulators (SL-MZM) with enhanced light-matter interaction for size and power reduction. 
It has a phase shifter length of 150$\sim$200 $\mu m$ and a footprint about 10$\times$ greater than Si MRR while an order of magnitude smaller than the typical foundry offered Si Mach-Zehnder modulator (MZM) PDK elements. 
This SL-MZM is thermally robust, with no thermal tuning/locking circuit needed, and can also tolerate large manufacturing variations.  
Different from a multi-wavelength dynamic PTC designs~\cite{NP_arXiv2023_Zhu}, \ours simplifies the spectral multi-wavelength encoding to high-speed temporal encoding, eliminating the need for complex dispersion-engineered broadband device designs such as Si modulators, optical power splitters and directional couplers as well as remove WDM MUX/DEMUX overhead. 

The major contributions of this paper are as follows:
\begin{itemize}
    \item We present a compact and energy-efficient multi-core photonic edge AI accelerator, \ours, with device and architecture co-optimization and customization.
    \item \textbf{Compact \& Efficient Photonic Components} -- 
    To enable ultra-fast, compact, low-power input operand encoding and dot-product computing, we adopt a customized slow-light MZM device with orders-of-magnitude smaller footprint and switching energy than the PDK MZM.
    We also customize optical power splitters with varying splitting ratios and an ultra-low power $\pi$/2 phase shifter.
    With customized devices, \ours is 6.8$\times$ more compact and 9.1$\times$ more power efficient than the foundry counterparts. 
    \item \textbf{Hierarchical Product Accumulation} -- \ours leverages photocurrent aggregation and temporal integration for partial product accumulation in the analog domain, significantly reducing the laser power and analog-to-digital conversion cost. 
    We also enable input modulator sharing and output readout circuitry sharing to minimize the E-O/O-E cost.
    \item \textbf{Versatile and Robust Edge AI Evaluation} --
    We evaluate \ours on both convolutional NNs and Vision Transformers on AR/VR speech recognition, image classification, and advanced semantic segmentation tasks and show comparable accuracy and superior robustness to low-bit quantization and hardware noises from experimental measurement.
    \item \textbf{New Area-Energy Efficiency Pareto Frontier} --
    We comprehensively evaluate the scalability and efficiency of our proposed \ours architecture and show 368.6 TOPS peak performance, 22.3 TOPS/W energy efficiency, and 1.2 TOPS/mm\textsuperscript{2} compute density, outperforming state-of-the-art electronic counterparts.
\end{itemize}

\section{Overview of Time-Multiplexed Dynamic PTC Architecture Design of \ours}

Matrix multiplication is the key linear operation for various information processing workloads. The proposed dynamic photonic tensor core will perform matrix-matrix multiplication. 
For generality, we consider two input matrices, matrix $X$ with $M\times N$ dimension and matrix $Y$ with $N\times Q$ dimension: 
\begin{equation}
    \label{eq:Matrix}
    X=\begin{bmatrix}
        x_{11} & \cdots & x_{1N} \\
        \vdots & \ddots & \vdots \\
        x_{M1} & \cdots & x_{MN}
    \end{bmatrix}, \quad
    Y=\begin{bmatrix}
        y_{11} & \cdots & y_{1Q} \\
        \vdots & \ddots & \vdots \\
        y_{N1} & \cdots & y_{NQ}
    \end{bmatrix}.
\end{equation}
The matrix multiplication $Z=X \cdot Y$ is
\begin{equation}
    \label{eq:DotProduct}
    \small
    Z=\begin{bmatrix}
        z_{11} & \cdots & z_{1Q} \\
        \vdots & \ddots & \vdots \\
        z_{M1} & \cdots & z_{MQ}
    \end{bmatrix}=\begin{bmatrix}
        x_{11} & \cdots & x_{1N} \\
        \vdots & \ddots & \vdots \\
        x_{M1} & \cdots & x_{MN}
    \end{bmatrix}\cdot\begin{bmatrix}
        y_{11} & \cdots & y_{1Q} \\
        \vdots & \ddots & \vdots \\
        y_{N1} & \cdots & y_{NQ}
    \end{bmatrix}.
\end{equation}
The resulting $Z$ is an $M\times Q$ matrix; and its $a$-th row, $b$-th column element $z_{ab}$ is obtained by calculating the dot-product of $a$-th row vector of $X$ and $b$-th column vector of $Y$, i.e., 
\begin{equation}
    \label{eq:VectorDotproduct}
    z_{ab}=X_a\cdot Y_b=\begin{bmatrix}
        x_{a1} & \cdots & x_{aN}
    \end{bmatrix}\cdot\begin{bmatrix}
        y_{1b}\\
        \vdots\\
        y_{Nb}
    \end{bmatrix}.
\end{equation}
Each vector dot-product operation can be mapped to a dynamic dot-product engine.
Multiple dot-product engines can form an array structure, i.e., a tensor core, to realize parallel matrix-matrix multiplication.
The design of the dot product engine will be discussed in Section~\ref{sec:DotProductEngine}, and the proposed PTC architecture will be explained in Section~\ref{sec:PTCArch}.

\subsection{Dynamic Photonic Dot-Product Engine}
\label{sec:DotProductEngine}
\begin{figure}[hbt!]
    \centering
    \includegraphics[width=1.1\columnwidth]{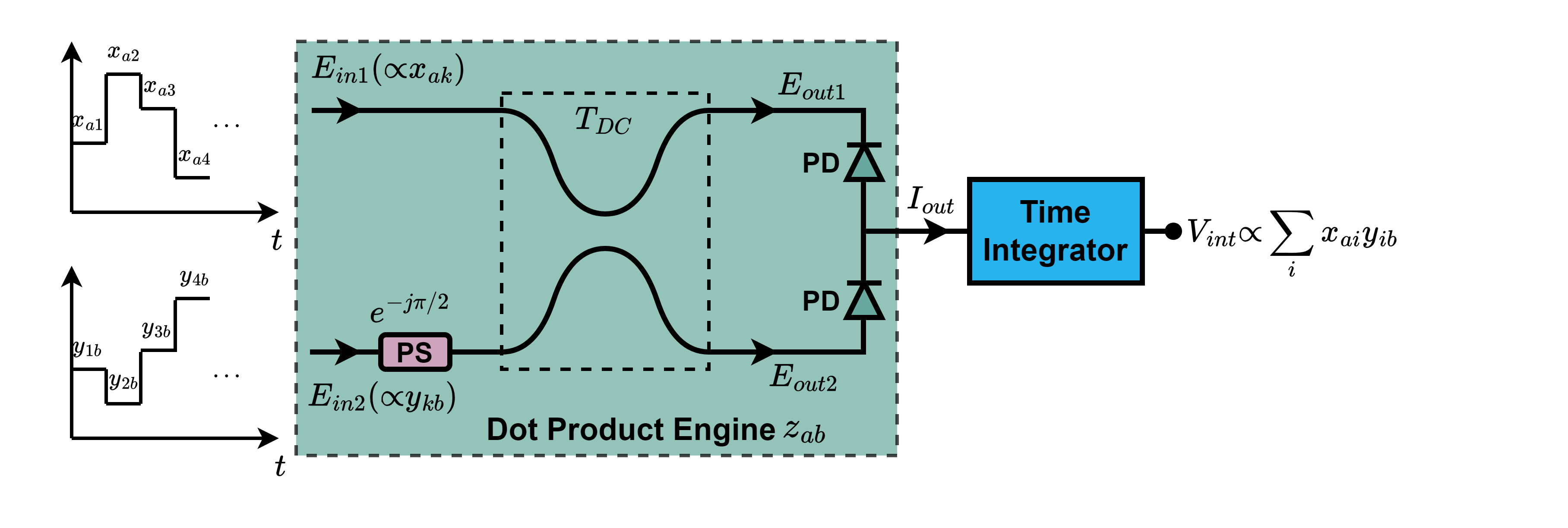}
    \caption{Schematic of a dynamic optical dot-product engine.
    }
    \label{fig:DotProductEngine}
\end{figure}

The matrix dot product operation that can be realized in photonic/electronic hardware is shown in Fig.~\ref{fig:DotProductEngine}. Matrix dot product calculates the element $z_{ab}$ while the data pairs ($x_{ak}$, $y_{kb}$) ($k=1, 2, ..., N$) are encoded to the phase and amplitude of input light to the directional coupler. 
A phase shifter (PS) is implemented in one input arm of the directional coupler to generate a $-\pi/2$ phase shift.
The core of the dot product engine consists of a 2$\times$2 directional coupler connecting followed by a pair of balanced photodetectors. 
The 2$\times$2 directional coupler provides interference between coherent light inputs of two arms. 
The transfer matrix for this structure with an ideal, lossless directional coupler can be expressed as 
\begin{equation}
    \label{eq:TranferMatrix}
    T_{DC}\cdot T_{PS}=\begin{bmatrix}
        t & j\kappa \\
        j\kappa & t
    \end{bmatrix}\cdot
    \begin{bmatrix}
        1 & 0 \\
        0 & e^{-j\frac{\pi}{2}}
    \end{bmatrix},
\end{equation}
where $t$ is the through-coupling coefficient, $k$ is the cross-coupling coefficient and $j$ is the imaginary unit. 
For dot product computing, 50:50-splitting is used, so $t=\kappa=\sqrt{2}/2$. 
Consider the electric fields of input signals to the directional coupler $[E_1, E_2]^T$ encoding a data pair $[x_{ak},y_{kb}]^T$, the output of the directional coupler $[E_{out1}, E_{out2}]^T$ can be expressed as
\begin{equation}
\small
    \label{eq:OutputExpression}
    \begin{aligned}
        \begin{bmatrix}
        E_{out1}\\
        E_{out2}
    \end{bmatrix}=T_{DC}\cdot T_{PS}\begin{bmatrix}
        E_1\\
        E_2
    \end{bmatrix}&=\frac{\sqrt{2}}{2}\begin{bmatrix}
        1 & j\\
        j & 1
    \end{bmatrix}\cdot
    \begin{bmatrix}
        1 & 0\\
        0 & -j
    \end{bmatrix}\cdot
    \begin{bmatrix}
        x_{ak}\\
        y_{kb}
    \end{bmatrix}\\
    &=\frac{\sqrt{2}}{2}\begin{bmatrix}
        x_{ak} + y_{kb}\\
        j(x_{ak}-y_{kb})
    \end{bmatrix}
    \end{aligned}.
\end{equation}
The photocurrent of the PDs connected to the directional coupler is proportional to the received optical power. Assume identical responsivity of two cascaded PDs, the output current $I_{out}$ can be calculated by
\begin{equation}
\small
    \label{eq:Photocurrent}
    I_{out}\propto |E_{out1}|^2-|E_{out2}|^2\propto |x_{ak}+y_{kb}|^2-|x_{ak}-y_{kb}|^2\propto x_{ak}y_{kb}.
\end{equation}
This is the product between two elements. 
To accomplish the dot-product operation between vector $X_a$ and $Y_b$, $x_{ak}y_{kb}$ needs to be summed up over all the $k$ labels from 1 to $N$.
The electrical modulated signal to the slow-light MZM follows the sample-and-hold operation to inject the vector elements {$x_{a1},x_{a2},\cdots,x_{aN}$} and {$y_{1b},y_{2b},\cdots,y_{Nb}$} through two slow-light MZMs sequentially. 
A time integrator is connected right after the dot product engine to operate the summation $\sum_{k=1}^N x_{ak}y_{kb}$ in the time domain so that the integrator readout voltage $V_{int}$ will represent the dot product between vector $X_a$ and $Y_b$,
\begin{equation}
    \label{eq:Integration}
    V_{int}\propto \sum_{k=1}^N x_{ak}y_{kb}\propto X_a\cdot Y_b.
\end{equation}
The detailed physical realization of the dot-product engine and time integrator will be discussed in Section~\ref{sec:Component}. 

\subsection{\ours Architecture Overview}
\label{sec:PTCArch}
\begin{figure*}
    \centering
    \includegraphics[width=0.9\textwidth]{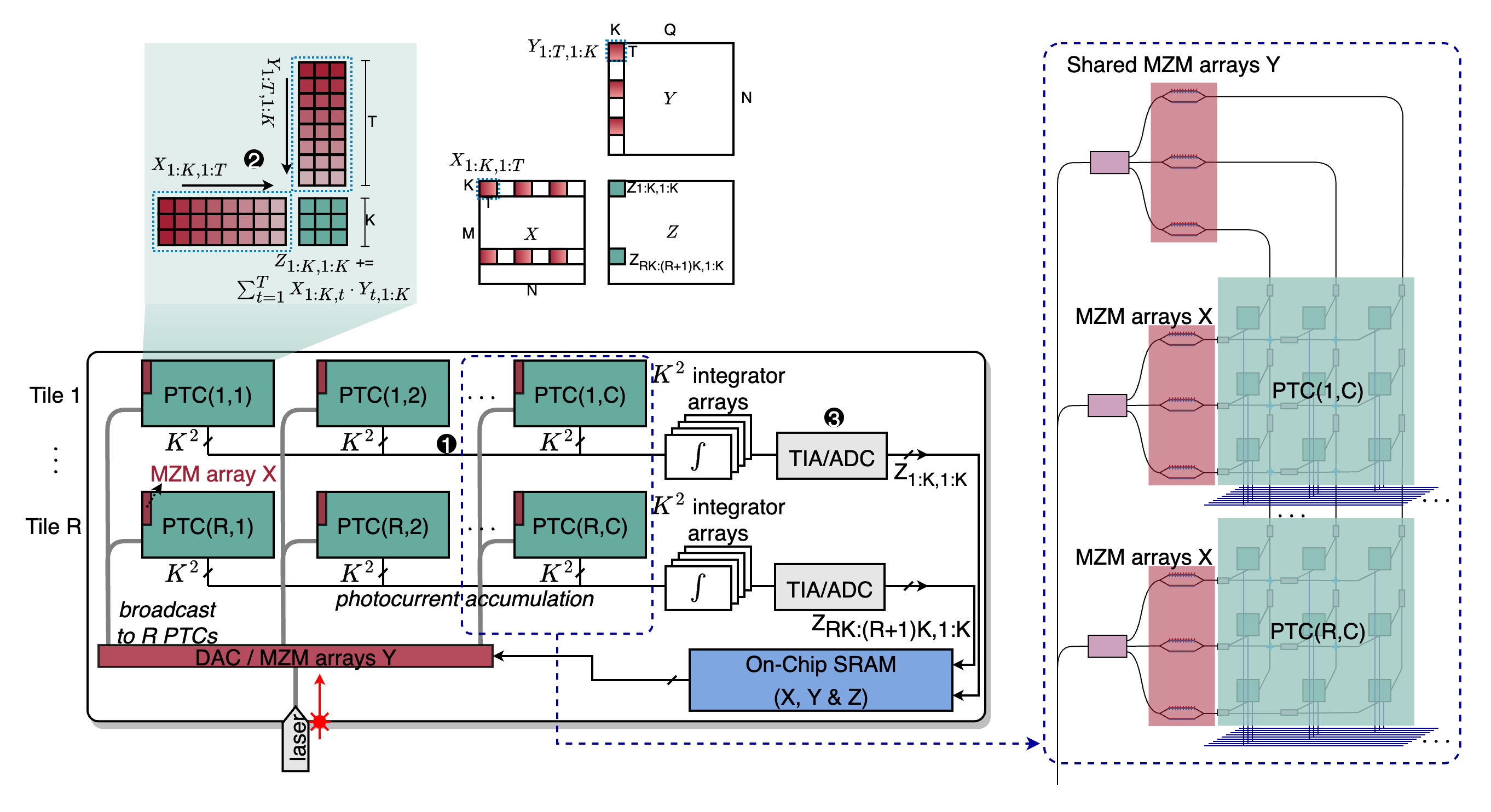}
    \vspace{-10pt}
    \caption{Our designed multi-core time-multiplexed dynamic photonic tensor accelerator \ours.
    \ding{202}-\ding{204} correspond to the hierarchical partial product accumulation in Eq.~\eqref{eq:PartialProduct}.
    All $R$ PTCs in a column share the same $Y$ matrix MZMs.
    All $C$ PTCs in a row share the same readout circuitry.
   }
    \label{fig:ArchOverview}
\end{figure*}

We have introduced one dynamic dot-product engine to realize vector dot-product.
Now, we introduce a multi-core time-multiplexed photonic tensor accelerator \ours for parallel dot-product, shown in Fig.~\ref{fig:ArchOverview}.

We have $R$ tiles in the architecture, and each tile contains $C$ PTCs.
Each PTC is a crossbar of $K\times K$ dynamic dot-product engines, which can finish a $K\times 1$ times $1 \times K$ vector outer product at each timestep.

Given an $M\times N$ times $N\times Q$ GEMM workload, we first partition the matrix $X$ into $M/K$ horizontal strips, each with a size of $K\times N$, and matrix $Y$ into $Q/K$ vertical strips, each with a size of $N\times K$.
One $K\times K$ block in the result matrix $Z_{1:K,1:K}$ can be computed by accumulating $N$ vector outer product, i.e., $Z_{1:K,1:K}=\sum_{t=1}^N~X_{1:K,t}\cdot Y_{t,1:K}$.
This length-$N$ reduction can be mapped to $C$ PTCs in a tile in parallel, and each PTC is responsible for computing $P=\frac{N}{C}$ vector outer products, which is formally rewritten as $Z_{1:K,1:K}=\sum_{p=1}^{P}(\sum_{c=1}^C~X_{1:K,(c-1)P+p}\cdot Y_{(c-1)P+p,1:K})$.
Therefore, the total cycles consumed to compute $Z_{1:K,1:K}$ is $P=\frac{N}{C}$.
There are $\frac{M}{K}\times \frac{Q}{K}$ of such matrix blocks in $Z$, and we mapped them to $R$ tiles in parallel.
This entire matrix multiplication requires in total $\frac{MQN}{RCk^2}$ cycles.

Each cycle is defined as (1) feeding one vector into our PTC, (2) reading out the outer product results as photocurrent, (3) converting it to the electronic domain, and (4) accumulating partial product.
As we mentioned above, each PTC consumes $P=\frac{N}{C}$ cycles to finish one $K\times K$ block in the $Z$ matrix, which means a conventional architecture needs to convert the photocurrent as electronic digital signals through trans-impedance amplifier (TIA) and analog-to-digital converter (ADC) \emph{at every cycle for each PTC} and accumulate the result digitally with adders and registers.
With a high data rate, e.g., 5-10 GHz, the AD conversion and digital accumulation cost is non-trivial, becoming a bottleneck of the performance and efficiency as the ADC power is proportional to its sampling frequency.

\noindent\textbf{\underline{Hierarchical Product Accumulation}} -- To resolve the AD conversion efficiency bottleneck, we adopt hierarchical accumulation both spatially and temporally in the analog domain.
The dot-product result is rewritten as
\begin{equation}
    \label{eq:PartialProduct}
    \begin{aligned}
        \small
        Z_{1:K,1:K}=\!\!\!\underbrace{\sum_{p=1}^{P/T}}_{\text{\ding{204}}}\underbrace{\sum_{t=1}^T}_{\text{\ding{203}}}\underbrace{\sum_{c=1}^C}_{\text{\ding{202}}}\!\!X_{1:K,(c-1)P+(p-1)T+t} Y_{(c-1)P+(p-1)T+t,1:K}.
    \end{aligned}
\end{equation}
\ding{202} At each timestep $t$, the photocurrent carrying the partial product results will first be aggregated from all $C$ PTCs in parallel within the same tile via analog current summation, corresponding to the most-inner summation in Eq.~\eqref{eq:PartialProduct}.
\ding{203} Then, the aggregated photocurrents will be further accumulated over $T$ timesteps at the temporal integrator but still in the analog domain.
\ding{204} After every $T$ timesteps, the partial sum will be converted to the digital domain via the analog-to-digital converters (ADCs), and the integrators will be reset and prepared for the following $T$ cycles.
With this hierarchical accumulation mechanism, the ADC conversion is minimized to merely $P/T$ times per matrix block, leading to $T$ times lower AD conversion frequency and, thus, power consumption.

\noindent\textbf{\underline{Input/Output Hardware Sharing}} -- To maximize the hardware sharing of the multi-core accelerator, we explore both input and output sharing.
For input sharing, $R$ PTCs across different tiles within the same column will share the same $Y$ vectors. 
Thus, the input vectors $Y$ can be modulated in the shared MZM arrays and broadcast to them via on-chip waveguide interconnects.
For output sharing, the partial products from $C$ PTCs within a tile are aggregated by summing up their photocurrent. 
For each tile, all $C$ PTCs share the same group of integrators, TIAs, and ADCs.
The total cost of those readout circuitry can be reduced by $C$ times with output sharing.

Next, we focus on the detailed design of a $K\times K$ time-multiplexed PTC to explain how our architecture performs dynamic matrix-matrix multiplication. 
For illustration simplicity, we set the matrix with an equal number of rows and columns, i.e., $K$, while the architecture can be applied to a matrix with arbitrary dimensions. 
A coherent monochromatic light source is used as the input to the photonic tensor core units. 
The input light is first fanned out to $2K$ waveguides via a $1\times 2K$ splitter.
Next, a slow-light Mach-Zehnder modulator (SL-MZM) is connected in each waveguide arm, serving as the input operand modulator of the PTC. 
Digital electrical signals carrying the matrix information are converted to analog optical signals represented by the amplitude and phase before optical signals reach the dot-product engine for computing. 
Let $E_{in}$ be the electric field of the input light to the SL-MZM, and the electric field of MZM output can be expressed as $E_{in} \cos\theta$, allowing broadband mapping of both positive and negative values.
We consider two optical routing schemes for the PTC architecture in this work, namely a double-layer-splitters scheme \oursd and an embedded-uneven-splitters scheme \ourse to guide the encoding optical signals to the targeting dot product engines. 
Schematics of the proposed PTC architecture are shown in Fig.~\ref{fig:DoubleLayerPTC} and Fig.~\ref{fig:UnevenPTC}. 

\subsubsection{Double-Layer-Splitter PTC Design \oursd}
\label{sec:DoubleLayerPTC}
\begin{figure}
    \centering
    \includegraphics[width=\columnwidth]{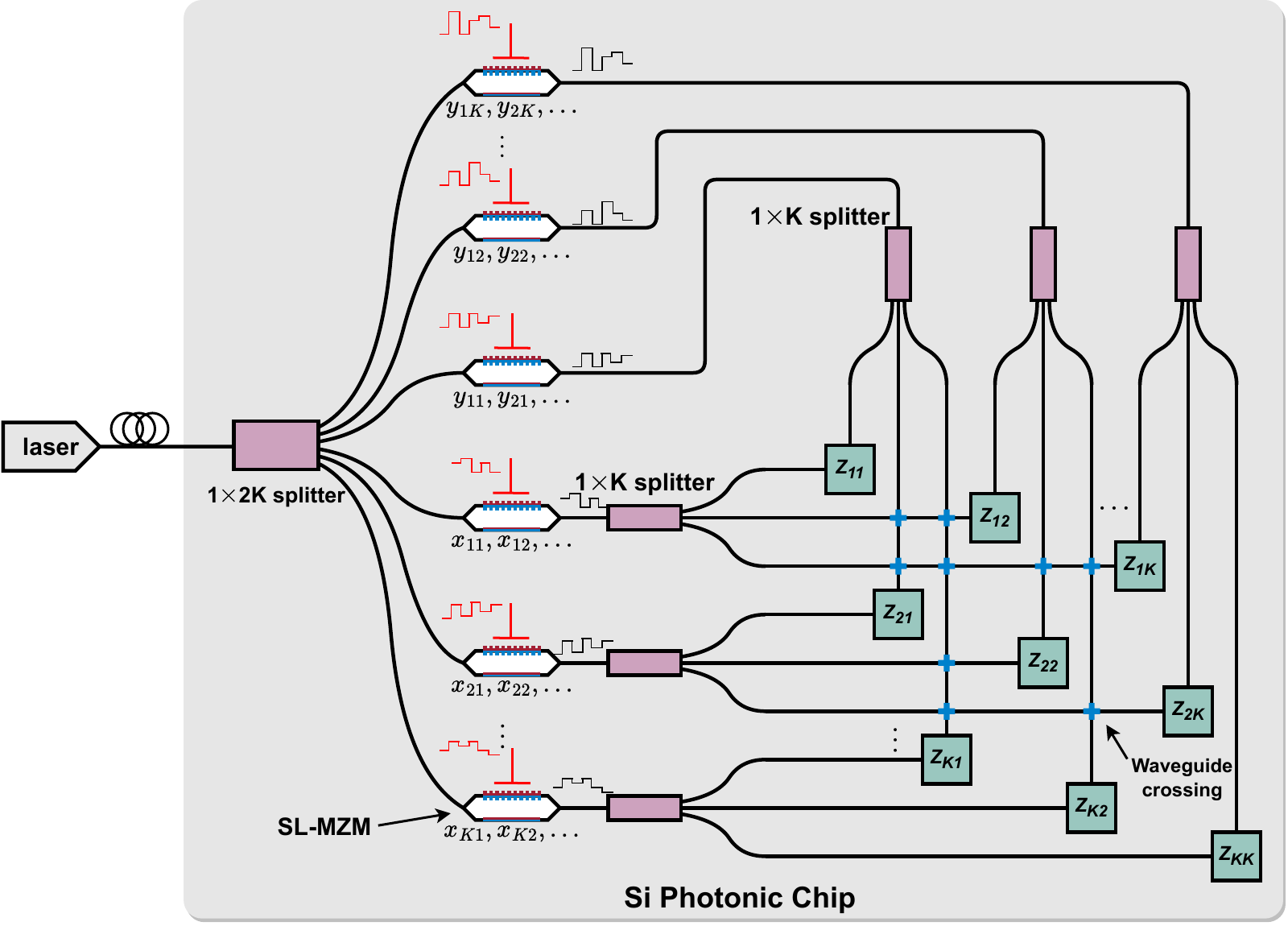}
    \caption{Schematic of our proposed time-multiplexed double-layer-splitter tensor core \oursd. $K=3$ is sketched here as an example for illustration.  }
    \label{fig:DoubleLayerPTC}
\end{figure}

A double-layer-splitter PTC design consists of two layers of optical splitters to route the encoded optical signals to the targeted dot product engines for matrix calculation, and a schematic of the architecture is shown in Fig.~\ref{fig:DoubleLayerPTC}. 
After the 1st fan-out $1\times 2K$ splitter, half of the optical paths (bottom $K$ paths) are used to encode matrix $X$ via an SL-MZM array, mapping to a row vector of matrix $X: X_a=[x_{a1},\cdots,x_{aK}], (a=1,2,\cdots)$. 
SL-MZMs on the top $K$ arms of the 1$\times2K$ splitter couple data of $N$ column vectors of matrix $Y: Y_b=[y_{1b},\cdots,y_{Kb}]^T, (b=1,2,\cdots)$. 
The second layer consists of $2K$ 1$\times K$ optical splitters, each of which evenly splits the optical power with encoded information into $K$ secondary output arms so that dot products between any pair of $X_a$ and $Y_b$ can be calculated simultaneously at $K^2$ dot product engines. 
Waveguide crossings are needed for this architecture. The coded optical signals may pass up to $(K-1)^2$ crossings to reach the dot product engine.  

\subsubsection{Embedded-Uneven-Splitters PTC Design \ourse}
\label{sec:UnevenPTC}
\begin{figure}
    \centering
    \includegraphics[width=\columnwidth]{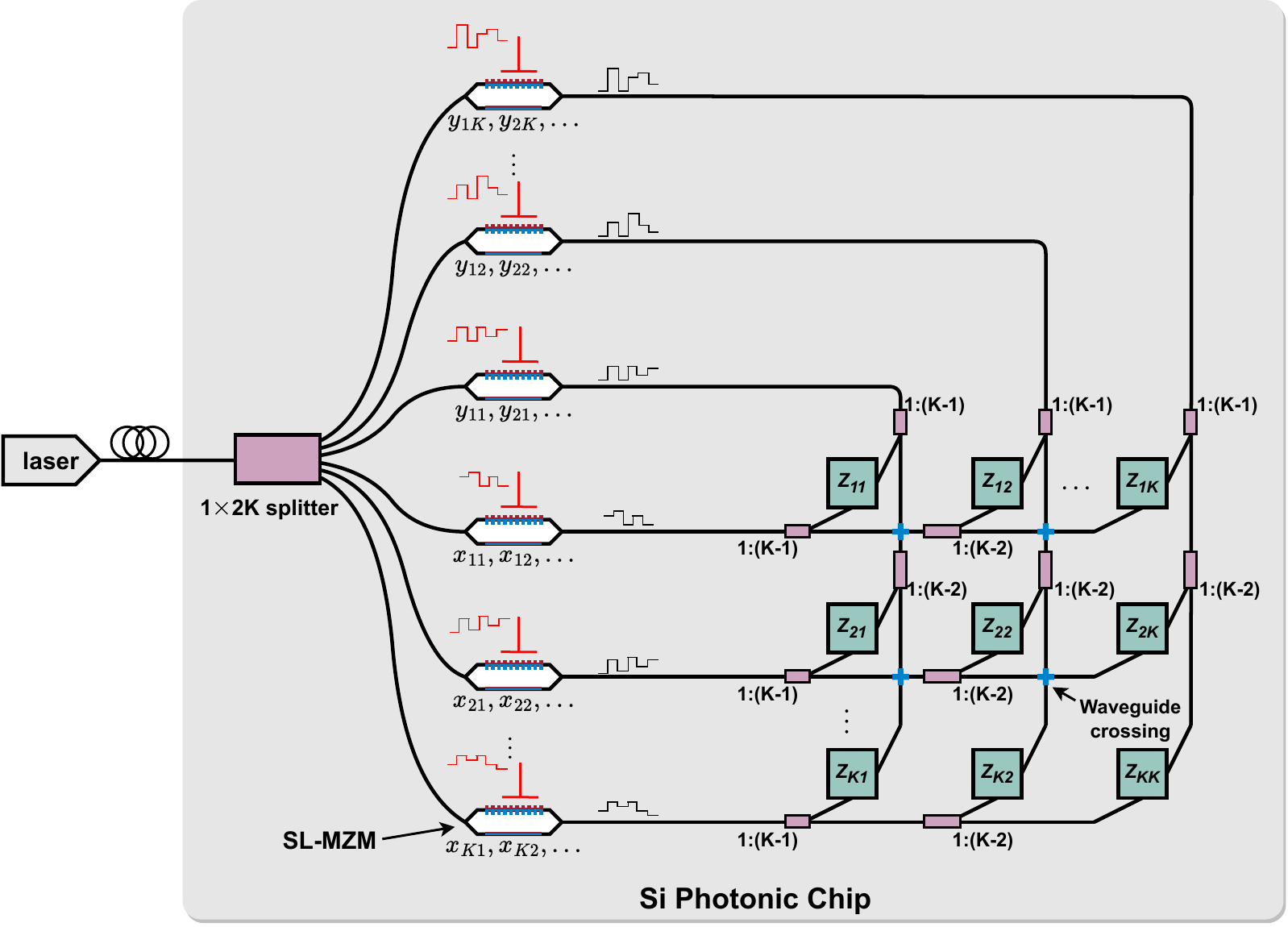}
    \caption{Schematic of our proposed time-multiplexed embedded-uneven-splitter tensor core \ourse. $K=3$ is sketched here as an example for illustration.}
    \label{fig:UnevenPTC}
\end{figure}

A schematic of the embedded-uneven-splitters PTC design, \ourse is illustrated in Fig.~\ref{fig:UnevenPTC}.
Different from the $\oursd$ design, this architecture adopts a series of uneven splitters to eliminate waveguide crossings.
The splitting ratios are set at $1:(K-1)$, $1:(K-2)$, $\cdots$, and $1:1$.
For a PTC with $K^2$ dot product engines, the splitting ratios of the two optical splitters that guide light into the dot product engine $z_{ab}$ are $1:(K-a)$ and $1:(K-b)$, respectively, to ensure identical input power to each dot product engine. 
The maximum number of crossings on the optical path is $K-1$.

Comparing \oursd with \ourse, \oursd design only requires 1 optical splitter before reaching the DOT engine with the cost of the increased number of waveguide crossings in some waveguide paths.
For the $\ourse$ design, the number of uneven power splitters and waveguide crossings needed in each path are both $K-1$, while \oursd design requires  $(K-1)^2$ waveguide crossings. We anticipate lower accumulated device loss in the \ourse design when  $K$ is large.  
In the following discussion, we only focus on the embedded-uneven-splitters design \ourse and simplify it as \ours.

\section{Photonic Components for PTCs}
\label{sec:Component}

\subsection{Laser Source}
\label{sec:Laser}
A PTC utilizing optical wave phase and amplitude in time-domain processing only requires a monochromatic light source for optical signal processing. 
In the realm of integrated photonic computing chip design, o-band operation, in comparison with c-band components, offers several distinct advantages such as a smaller optical mode volume in Si/SiO$_\text{2}$ waveguide structure, higher mode confinement with tighter bending radius and > 1.5$\times$ higher in Ge PD responsivity ~\cite{NP_eng2015state, NP_zhao2021high}.
 
The second consideration pertains to the choice between an on-chip III-V integrated laser diode and an off-chip laser module. 
While the heterogeneously bonded laser to Si holds the promise of the miniaturized, photolithographically defined coherent on-chip light source, it has yet to mature for mass production. 
The long-term reliability of on-chip lasers remains undetermined. 
Laser cavities are highly sensitive to temperature variations, thus heterogeneously intergated on-chip laser, being in the close vicinity of other electronics that generate considerable heat would demand more complex electronics circuits in thermal management to maintain on-chip laser diode emission stability in optical mode/wavelength, polarization, and optical power. 
Integrated optical isolators on Si platform are not yet available from SiPho foundry; while an optical isolator is critical in minimizing reflections that could disturb laser operation if the reflection is not addressed. 
Varying laser operation will also, in turn, degrade the PTC performance. 
In this work, we advocate a technological path that utilizes a separate, off-chip laser module that takes advantage of the latest advancement in optical packaging to achieve low insertion loss at the fiber-to-chip interface. 
 
High-power monolithic o-band lasers, capable of producing output powers as high as 150 mW \cite{NP_optilab}, are commercially available now. 
In this work, we utilize a moderate laser power of 100 mW for system power-related analysis and evaluation. 
Utilizing index-matched epoxy and emerging packaging technology, such as photonic wires~\cite{NP_packaging_barwicz2018automated, NP_packaging_khan2020low, NP_packaging_nauriyal2019fiber}, one can expect 0.5dB - 2dB insertion loss at the fiber to chip facet.

\subsection{Slow-Light Mach-Zehnder Modulator}
\label{sec:SLMZM}
\begin{figure}[hbt!]
    \centering
    \subfloat[]{\includegraphics[width=0.495\columnwidth]{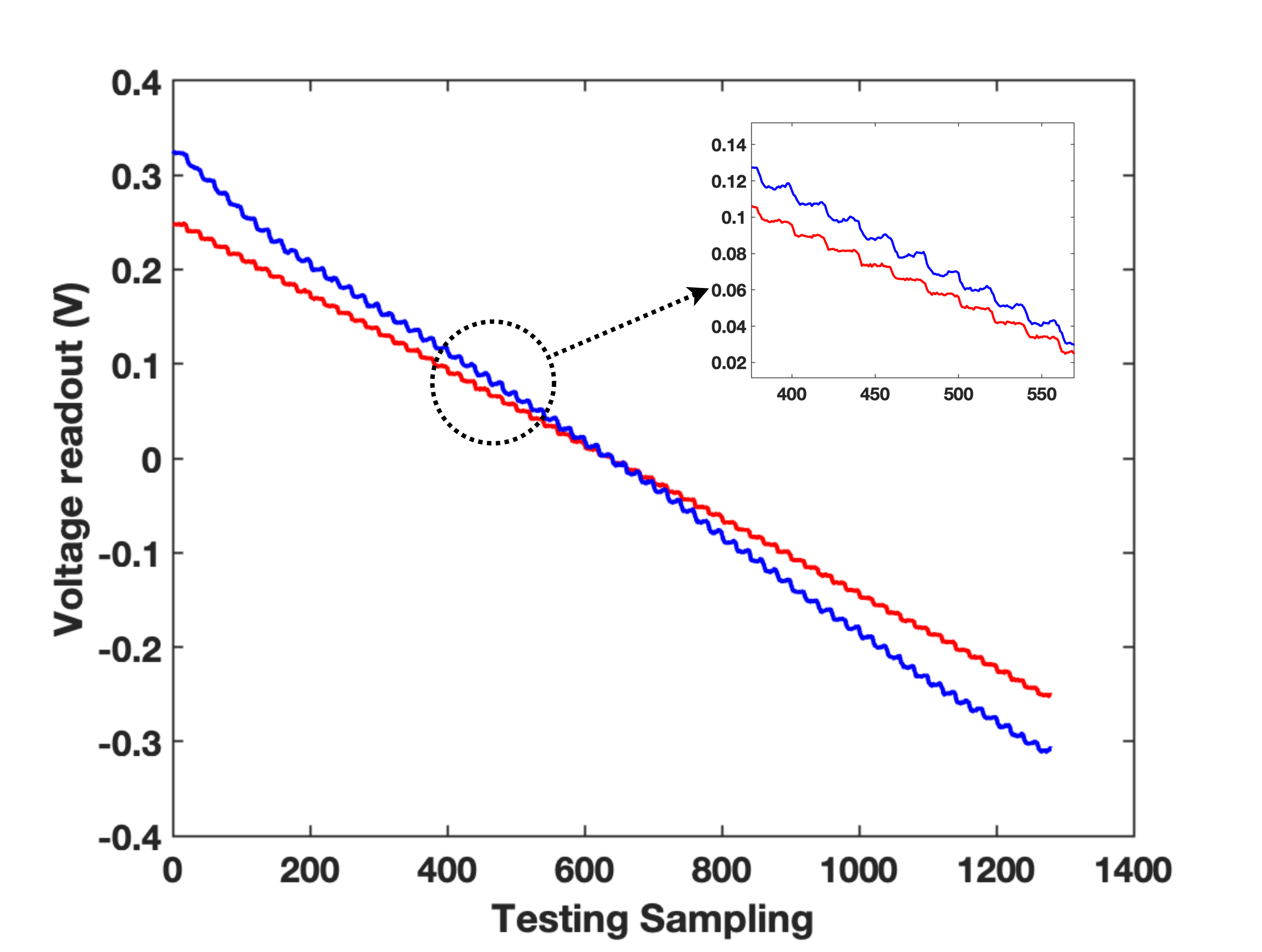}
    }
    \subfloat[]{\includegraphics[width=0.495\columnwidth]{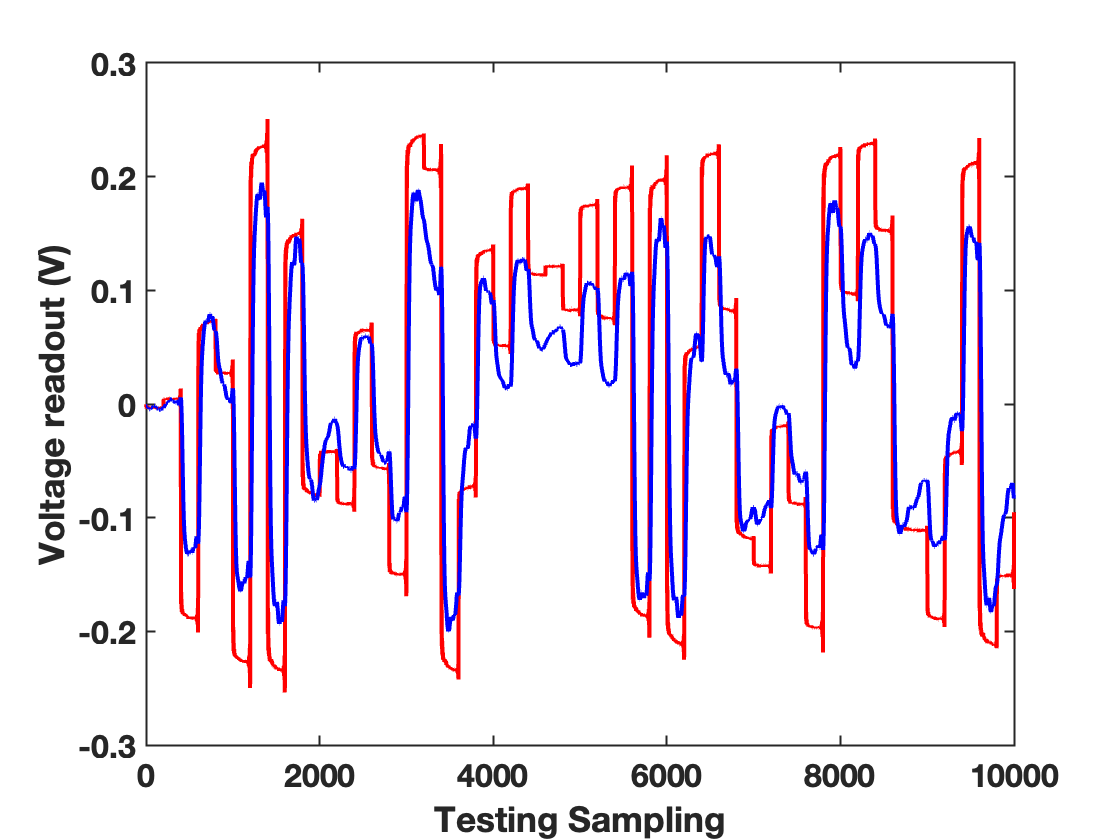}
    }\\
    
    \caption{Bit resolution testing of SL-MZM at 100 MHz clock frequency with (a) 6-bits staircase signal and (b) 6-bits random signal. The red curves show direct driving signals from the arbitrary waveform generator (AWG), while the blue curves represent the SL-MZM response readout by the on-chip PD.}
    \label{fig:MZMTestResult}
\end{figure}

Mach-Zehnder Modulators (MZMs) play a crucial role in the conversion of electrical signals to the optical domain in chip-scale PTC. Si modulators, utilizing the carrier plasma effect, offer a cost-effective and high-density integration solution for on-chip PTC. 
Achieving
a dot product operation for matrices of the size of $K\times K$ requires 2$K$ modulators for signal conversion.  
The physical dimension of these Si modulators serves as a critical design parameter, impacting the scalability of matrix operation. 
In this study, our approach involves the adoption of a 1D dielectric photonic crystal waveguide, specifically a rectangular-shaped Bragg grating~\cite{NP_IEEE_chen2022}, as a slow-light-enabled compact modulator to significantly reduce the footprint of the modulator array~\cite{NP_IEEEPTL2023_Anderson, NP_IEEE_anderson2022}. 
Lately, we have experimentally demonstrated a Si slow-light MZM (SL-MZM) with a phase shifter length ($L_{PS}$) of 150 $\mu m$ for optical compute application~\cite{NP_CLEO2024_ours}. 
The SL-MZM reported in this work was fabricated at AIM Photonics under a multi-project wafer (MPW) run, ensuring complete foundry compatibility.  
The modulator output is routed to an on-chip Ge photodetector (PD), a standard AIM PDK component with a tested bit rate of 15 Gbps. 
The SL-MZM, operating under maximum $V_{pp}$ signals of 3.5V, was characterized with up to 6-bit of resolution using both staircase and random data inputs. The readout signals from the PD are displaced on a real-time oscilloscope, shown in Fig.~\ref{fig:MZMTestResult}. 
The averaged variance during bit-holding time is reported as $9.72\times 10^{-7}$ and $6.59\times 10^{-5}$ for the staircase and random signal input cases, respectively. 

Reflection occurring at different junctions within the modulator device, optical absorption due to carriers in waveguides, propagation loss in the Bragg grating phase shifter due to increased group indices, and mode mismatch at the Bragg grating waveguide interfaces are the primary factors contributing to the modulator insertion loss. The measured total modulator insertion loss is $\sim$6.4 dB for $L_{PS}$ = 150 $\mu m$ and is utilized as the loss figure in the system evaluation.  

To achieve high-bit resolution at a high computing clock frequency, it is imperative to optimize both the electrical bandwidth and linearity of a Si modulator. Operating under reverse bias, the speed of a Si SL-MZM is limited by its RC time constant and photon lifetime. Typically, the PN junctions are doped at an elevated level (ranging from $10^{18}/cm^3$ to $10^{19}/cm^3$) to enhance the carrier plasma effect. As the phase shifter length is reduced in an SL-MZM, the total capacitance decreases. In this work, the measured SL-MZM junction capacitance $C_j$ was approximately $\sim$0.75 pF. Depending on the doping level in the connecting Si bar from the ridge waveguide to the via contacts, the intrinsic resistance of a SL-MZM ranges from 5 to 10 ohms. The estimated RC time-limited electrical bandwidth of a SL-MZM is thus in the hundreds of GHz. The slow-light effect can be viewed as a traveling wave resonant in its propagation direction, with the optical bandwidth determined by the Q-factor of the resonator. For the rectangular Bragg grating-shaped slow-light, an optical bandwidth of approximately $\sim$26 GHz is estimated~\cite{NP_IEEEPTL2015_Caverley}. However, the SL-MZM of this work didn't reach its maximum bandwidth potential due to impedance mismatch of the electrodes~\cite{NP_IEEEPTL2023_Anderson}, mismatch of the RF signals speed with the optical wave with high group index~\cite{NP_hinakura2018electro} and waveguide dispersion in the slow light spectrum. Dispersion engineering techniques such as phase-shifted Bragg grating, dispersion compensation~\cite{NP_khan2012}, and line-shift photonic crystal waveguide are all effective approaches in reducing the dispersion-induced bandwidth penalty. With careful device design and optimization, a SL-MZM operating at a 5GHz clock frequency is feasible, as assumed for system-level performance evaluation in this study. 

\subsection{Optical Power Splitter}
\label{sec:PowerSplitter}
The optical splitter is a crucial passive photonic component in integrated photonic systems for splitting optical power. 
Various types of structures such as Y-junction splitters~\cite{NP_Ysplitter_zhang2013compact}, multimode interferometers (MMIs)~\cite{NP_QISSC2019_VanNiekerk, NP_OFC2019_Liu} and directional couplers~\cite{NP_DC_sattari2020compact} have been demonstrated to achieve power splitting with varying splitting ratios. 
Y-junction splitters are usually compact and broadband, but the sharp corners can lead to increased reflection, resulting in unwanted FR resonance in a photonic system \cite{NP_Ysplitter_zhang2013compact}. The MMI-based power splitter is suitable for $1\times K$ uniform power splitting, while the shape of tapered input and output waveguides needs to be carefully designed \cite{NP_MMI1xK_yao2021compact}. By adjusting the coupling length, a directional coupler can also be used to obtain varying optical power splitting. 

\subsubsection{$1\times 2K$ Optical Splitter}
\label{sec:Splitter2K}
\begin{figure*}[hbt!]
    \centering
    \subfloat[]{\includegraphics[width=0.3\textwidth]{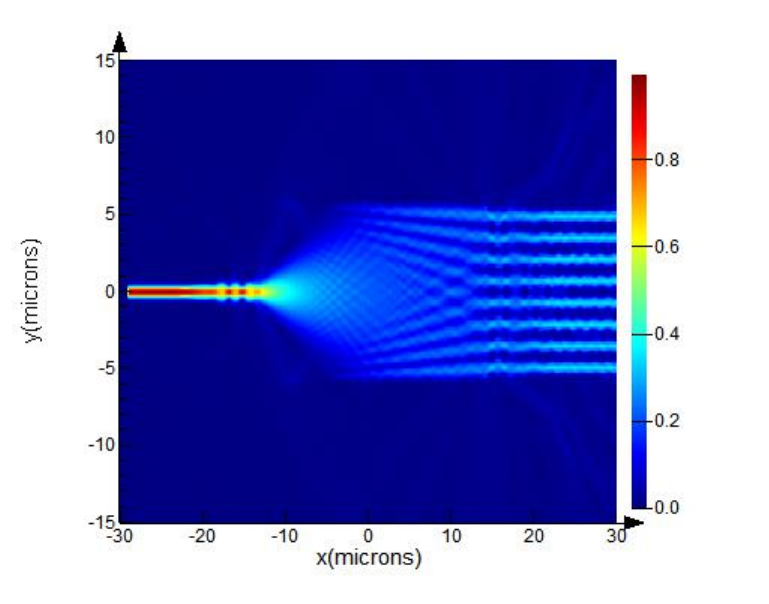}
    }
    \subfloat[]{\includegraphics[width=0.3\textwidth]{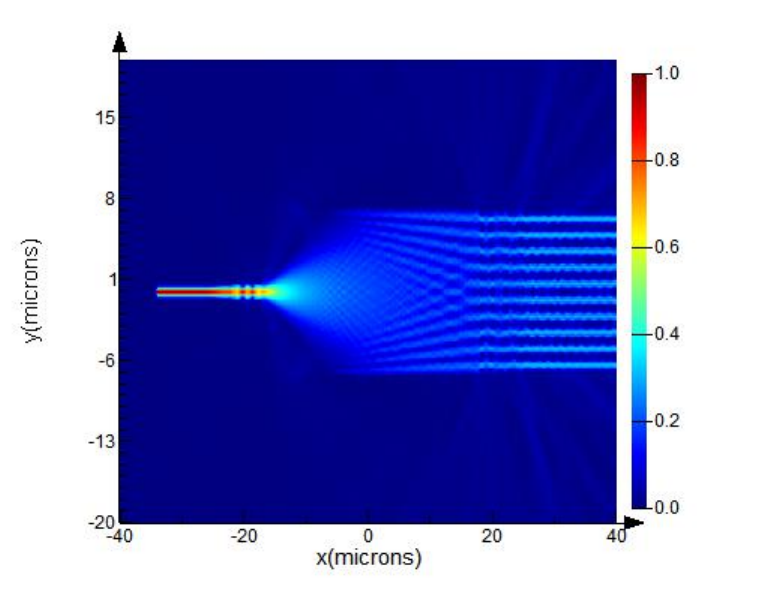}
    }
    \subfloat[]{\includegraphics[width=0.3\textwidth]{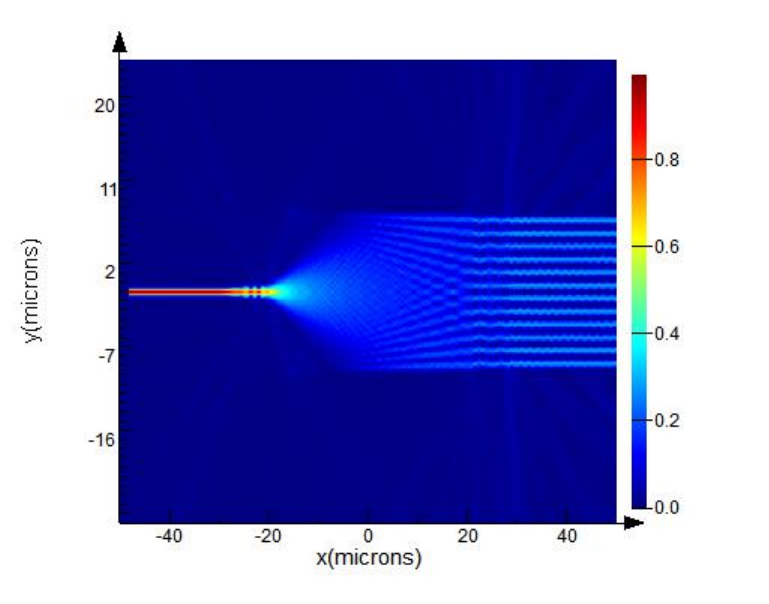}
    }
    \caption{Simulated electric field profiles of (a) 1$\times$8, (b) 1$\times$10, and (c) 1$\times$12 MMIs.}
    \label{fig:MMISimulationResult}
\end{figure*}

In our proposed PTC, the first layer $1\times 2K$ splitter adopts the design of MMI to fan out the CW laser light to $2K$ slow-light MZMs. 
For a center-excited $1\times K$ MMI splitter, $K$-folded self-imaging can be reproduced at MMI output when the length of the multimode waveguide section $L_{MMI}$ satisfies
\begin{equation}
    \label{eq:MMILength}
    L_{MMI}=\frac{3iL_{\pi}}{4K}, i=1,2,3,\cdots,
\end{equation}
where $L_{\pi}$ represents the beating length given by
\begin{equation}
    \label{eq:LPI}
    L_{\pi}=\frac{\pi}{\beta_0 - \beta_1}\approx \frac{4n_{eff}w_e^2}{3\lambda_0}.
\end{equation}
Here $\beta_0$, $\beta_1$ represent the propagation constants of the fundamental mode and first-order mode, $n_{eff}$ is the effective index of the multimode waveguide section, $\lambda_0$ is the operated wavelength and $w_e$ is the effective width and can be approximated as the multimode waveguide section width $W_{MMI}$ in silicon photonics \cite{NP_JLT1995_Soldano}. 
We use the 1$\times$8 MMI design in \cite{NP_MMI1xK_yao2021compact} for $K=4$ case and develop the 1$\times$10 and 1$\times$12 MMI designs based on Eq.~\eqref{eq:MMILength}. 
Consider a silicon waveguide layer thickness of 220 nm. The waveguide width is  450 nm and tapered to 1.2 $\mu m$ at the multimode waveguide segment.  
The simulated electric field profiles of 1$\times$8, 1$\times$10, and 1$\times$12 MMIs are shown in Fig.~\ref{fig:MMISimulationResult}, corresponding to $N=4$, $N=5$, and $N=6$ scenarios. The dimensions ($L_{MMI}\times W_{MMI}$) of 1$\times$8, 1$\times$10 and 1$\times$12 MMIs are 27.8 $\mu m$$\times$11.3 $\mu m$, 34.6 $\mu m$$\times$14.1 $\mu m$, 41.4 $\mu m$$\times$16.9 $\mu m$, respectively. 
The insertion loss (IL) is calculated to be 0.14dB, 0.20dB, and 0.21dB for 1$\times$8, 1$\times$10, and 1$\times$12 MMIs, respectively.
We adopt the 1$\times$10 MMI as a base design and assume a linear scaling law in MMI's length/width for a generic $1 \times 2K$ MMI and a near-constant insertion loss regardless of fanout in later discussion.

\subsubsection{Optical Power Splitter Guiding to the Dot-Product Engine }
\label{sec:SplitterDC}
The \ours adopts directional couplers with varying splitting ratios to guide the coded optical signals to each DOT engine for matrix computing.  
A directional coupler with even splitting is often offered as a standard PDK component from SiPho foundries. Keeping the waveguide gap constant, one only needs to change the coupling length to adjust the splitting ratio. 
With 480 nm waveguide width and 200 nm gap between two parallel waveguides in the coupling region, our simulation shows that the coupling length is 14.6 $\mu m$, 11.2 $\mu m$, 9.2 $\mu m$, 8 $\mu m$, and 7 $\mu m$ to achieve splitting ratios of 1:1, 1:2, 1:3, 1:4, and 1:5, respectively.

\subsection{Dot Product Engine Design}

The dot product engine to realize vector-vector dot product is the key computation unit in our proposed photonic tensor core. 
A dot product engine consists of a 2$\times$2 optical power splitter, a $\pi/2$ phase shifter, a pair of balanced PD, and a time integrator.
They will be discussed separately in this section. 

\subsubsection{2$\times$2 Optical Power Splitter}

A 2$\times$2 50:50 optical power splitter is needed to generate interference between the optical signals from 2 input arms. 
both directional couplers and MMIs can be used to generate 50:50 power splitting. The directional coupler consists of two closely placed parallel waveguides, and the splitting ratio is wavelength-dependent, thus sensitive to the fabrication accuracy. 
The 2$\times$2 MMI power splitting is less wavelength sensitive than the directional coupler, while it is challenging to achieve an exact 50:50 splitting ratio, and insertion loss is usually higher than the directional coupler. 
Two interference mechanisms, namely paired interference and general interference, can be applied to MMI design. 
The paired interference mechanism is generally used for designing $2\times K$ MMIs, where the modes contributing to the imaging in the multimode section are paired \cite{NP_JLT1995_Soldano}. 
The length of the multimode waveguide section $L_{MMI}$ satisfies 
\begin{equation}
    \label{MMIlength_paired}
    L_{MMI}=\frac{iL_{\pi}}{K}, i=1,2,3,\cdots.
\end{equation}
The two input waveguides have to be placed at $+\frac{W_{MMI}}{6}$ and $-\frac{W_{MMI}}{6}$ vertically from the center. 
For 2$\times$2 MMI based on a general $K\times K$ interference mechanism, there is no restriction on the location of the input waveguides \cite{NP_JLT1995_Soldano}. The length of the multimode waveguide section $L_{MMI}$ can be expressed as
\begin{equation}
    \label{MMIlength_general}
    L_{MMI}=\frac{3iL_{\pi}}{K}, i=1,2,3,\cdots.
\end{equation}
\begin{figure*}
    \centering
    \subfloat[]{\includegraphics[width=0.305\textwidth]{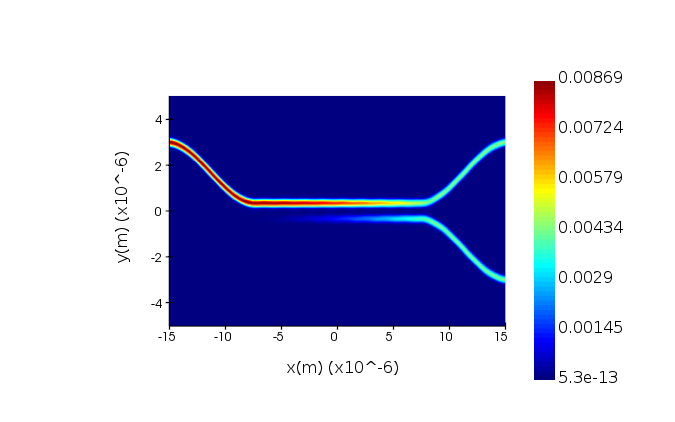}
    }
    \subfloat[]{\includegraphics[width=0.30\textwidth]{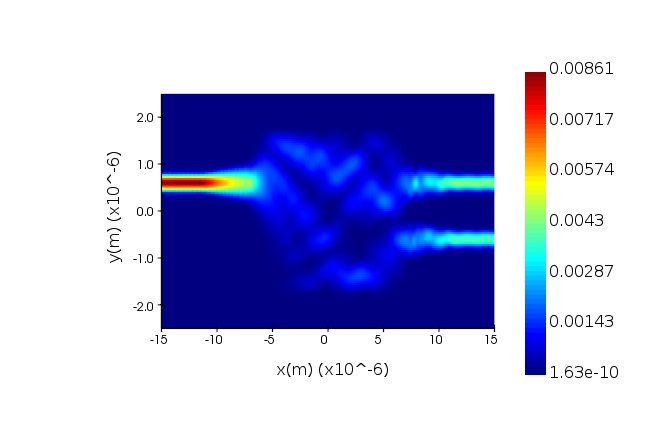}
    }
    \subfloat[]{\includegraphics[width=0.35\textwidth]{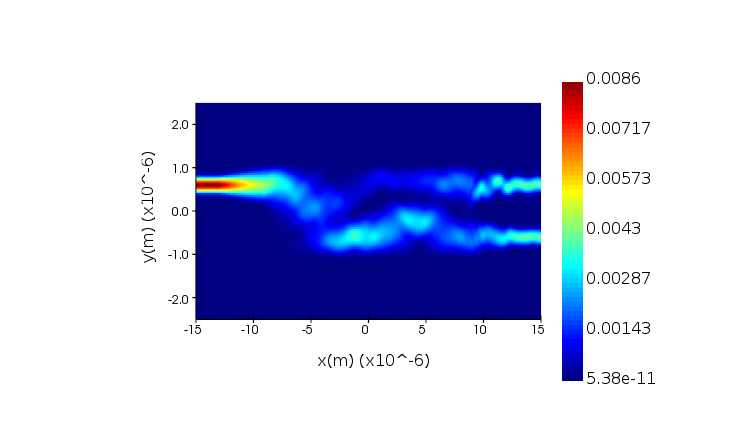}
    }
    \caption{Simulated electric power profiles of (a) directional coupler, (b) MMI with paired interference mechanism, and (c) MMI with general interference mechanism.}
    \label{fig:2x2CouplerSimulation}
\end{figure*}

\begin{table*}[hbt!]
\caption{Simulation results of three 2$\times$2 optical splitter designs.}
\label{tab:2x2powersplitter}
\resizebox{0.9\textwidth}{!}{
\begin{tabular}{|c|c|c|c|}
\hline
Splitter Design                         & Directional Coupler               & MMI (Paired Interference)     & MMI (General Interference) \\ \hline
Optical Coupling Region Dimension ($W\times L$)            & 1.2 $\mu m$ $\times$ 14.6 $\mu m$        & 3.6 $\mu m$ $\times$ 14.5 $\mu m$        & 2.2 $\mu m$ $\times$ 18 $\mu m$        \\ \hline
Block Dimension ($W\times L$) & 6.5 $\mu m$ $\times$ 31 $\mu m$        & 7 $\mu m$ $\times$ 40.5 $\mu m$       & 7 $\mu m$ $\times$ 44 $\mu m$      \\ \hline
Splitting Ratio  & 50:50        & 52.5:47.5      & 50:50         \\ \hline
Bandwidth at Targeting Splitting Ratio  & 1550 nm    & 1500 nm $\sim$ 1600 nm     & 1530 nm $\sim$ 1570 nm     \\ \hline
Insertion Loss at 1550 nm      & 0.05dB     & 0.18dB    & 0.37dB    \\ \hline
\end{tabular}
}
\vspace{-10pt}
\end{table*}

$L_{\pi}$ in Eq.~\eqref{MMIlength_paired} and Eq.~\eqref{MMIlength_general} follows the same definition as Eq.~\eqref{eq:LPI}.
Three 2$\times$2 optical power splitter designs are developed, and the results are summarized in Table~\ref{tab:2x2powersplitter}. 
The simulated electric field profiles are illustrated in Fig.~\ref{fig:2x2CouplerSimulation}, where the optical power is coupled in through one input arm, and output power is measured through both output arms. 
Overall, the directional coupler features lower insertion loss and smaller size, while the two MMI designs have larger bandwidth near the targeting 50:50 splitting ratio.
Taking the dimension, splitting ratio, and insertion loss into consideration, the directional coupler-based 2$\times$2 optical power splitter design will be utilized in the following system-level simulation study.

\subsubsection{ \texorpdfstring{$\pi/2$}{pi/2} Phase Shifter}
\begin{figure}[hbt!]
    \centering
    \includegraphics[width=0.5\columnwidth]{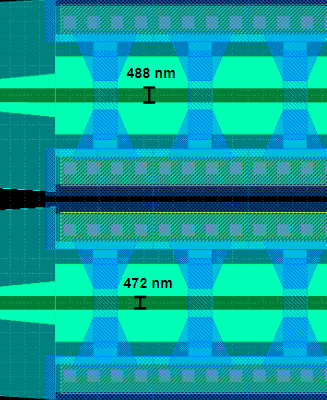}
    \vspace{-5pt}
    \caption{Side-by-side comparison of the thermo-optic phase shifters with a waveguide width difference.}
    \label{fig:phaseShifterWaveguideWidth}
\end{figure}

Maintaining a consistent $\pi/2$ phase difference between two optical paths can be realized through the utilization of either a path length difference or a waveguide effective index difference. In practice, there will be deviations from the targeted phase shifter (PS) owing to variations in waveguide dimensions induced during the manufacturing process. Thermal tuning is an effective method to adjust the offset to reach a precise $\pi/2$ phase difference. Optimized for the lowest static thermal tuning power, we adopt the design of $n_{\text{eff}}$ difference to achieve a $\pi/2$ phase shifter. The difference in phase $\phi$ between the two arms with identical lengths is 
\begin{equation}
    \label{eq:phaseShiftEquation}
    \beta_1L-\beta_2L = \phi,
\end{equation} where $\beta_1$ and $\beta_2$ represent the propagation constants of the two arms and $L$ is arm length. 
We set the global waveguide width to 480 nm while the two arms are set at 488 nm and 472 nm. 
A 5 $\mu$m taper is connected to the PS region. 
For a PS length of 30 $\mu$m, it will produce a $\sim\pi/2$ phase difference. 
A resistive heater is placed in the optical path of both arms following the design in \cite{NP_OptExpress2014_Harris}. 
As those two arms are placed in close vicinity, we anticipate minimum width difference variation, though their actual dimensions can deviate substantially from targeted values. 
In an extreme fabrication variation scenario of 488+2 nm and 472-2 nm, the phase difference is  0.6345 $\pi$, corresponding to an estimated heater tuning power of 5 mW to reach $\pi/2$.
When the fabrication variation is relatively small with advanced fabrication technology, we only need negligible active tuning power to compensate for the phase errors.

\subsubsection{Photodetector Responsivity and Sensitivity}
The sensitivity and responsivity of photodetectors are closely related to the laser power requirement and integrator designs.
Sensitivity $S_{PD}$ defines the minimum gap between two levels of optical power received by photodetectors given a certain bit error rate.
The loss of the circuit, including power splitting loss and insertion loss, is as follows
\begin{equation}
    \label{eq:IL}
    \begin{aligned}
        IL=&IL_{couple}+10\log_{10}K^2+IL_{MZM}+(K-1)IL_{cross}\\
        &+K\cdot IL_{splitter}+{IL}_{PS}+IL_{DC}.
    \end{aligned}
\end{equation}

Given the circuit insertion loss IL and PD sensitivity, we can derive the laser power (mW) requirement for each PTC to obtain $b$-bit output resolution,
\begin{equation}
    \label{eq:LaserPower}
    \frac{P_{laser}\cdot (1-10^{-ER/10})}{10^{IL/10}} \ge I_{noise}/R_{PD} + 2^b\cdot 10^{S_{PD}/10},
\end{equation}
where $I_{noise}$ is the dark current noise floor of the PD, $R_{PD}$ is the PD responsivity, and ER is the modulator extinction ratio.
$(1-10^{-ER/10})$ is the power penalty to compensate for the range reduction due to the non-ideal ER.
For example, with 20 dB insertion loss, 1 A/W responsivity, 20 nA dark current noise floor, 10 dB extinction ratio, and -27 dBm PD sensitivity, the minimum optical power from the laser to obtain 6-bit output is 14.2 mW.

Meanwhile, the balanced PD's current range determines the integrator's design.
Given the principle of time integration, i.e., $V_{out}\propto \int_t \frac{I_{PD}}{C_{int}} \text{dt}$, the maximum voltage with $T$-timestep integration of $f$-frequency datarate is $V_{max}\propto \int_0^{T/f} \frac{I_{PD,max}}{C_{int}} \text{dt}=T\cdot I_{PD,max}/(C_{int}f)$.
To avoid saturation-induced integration error, i.e., $V_{max} \leq V_{DD}$, we must carefully design the integration timestep $T$ and the capacitance $C_{int}$ given the maximum photocurrent generated by the balanced PD.
Detailed integrator design specifications are introduced in the following section.

\subsubsection{Temporal Integrator}
\label{sec:Integrator}

The proposed time-multiplexed approach requires integration of photodetector output current for the accumulation operation as in Eq.~\eqref{eq:Integration}. 
This is one of the key mechanisms in \ours to significantly relieve the ADC power bottleneck.

\begin{figure}[hbt!]
    \centering
    \includegraphics[width=\columnwidth]{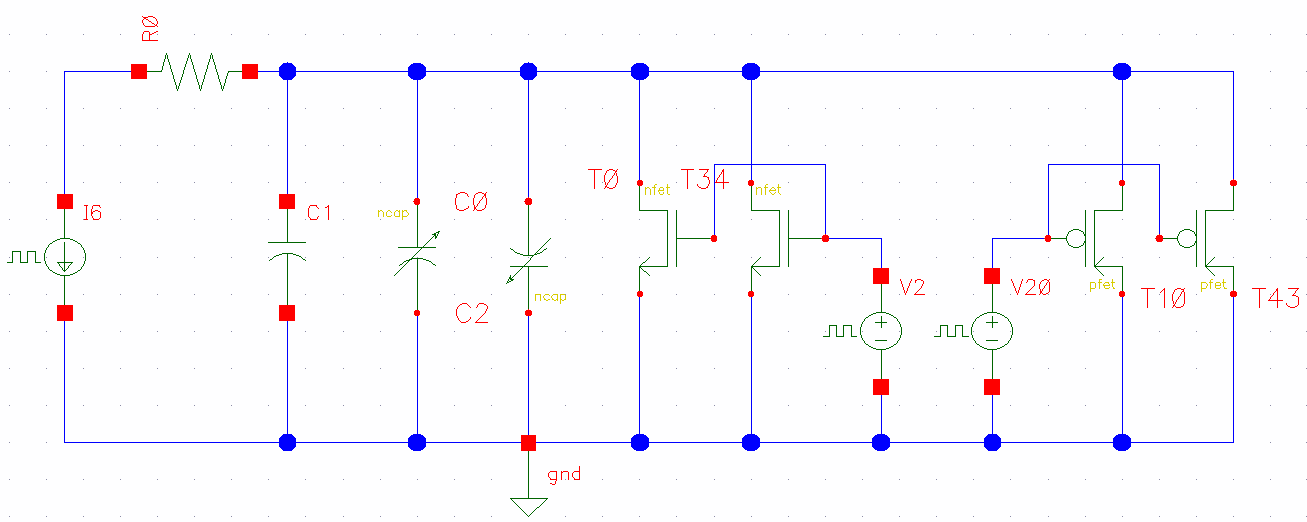}
    \caption{Schematic of the capacitive temporal integrator.}
    \label{fig:int_circuit}
\end{figure}

\noindent\textbf{\underline{Integrator Design and Optimization}} -- 
Our integrator design objective is to support a target maximum integration timestep $T$ with good linearity in the voltage response and fast reset speed.
We adopt a simple, compact, and foundry-compatible means of time integration using a capacitor. 
Capacitive elements are well suited for analog integration of current-based signals. 
The voltage across the terminals is proportional to the time-integral of the current from the photodiode.  
After each multiply-accumulate operation is complete, the capacitor integrator will need to be discharged (reset) before the next operation.  
By turning on field-effect transistors (FETs) in parallel to the capacitor, the charge across the capacitor can be rapidly dissipated for reset. 

Now, we show the detailed integrator design with a target maximum integration timestep $T$ and linearity and reset speed considerations.
The proposed integration unit is shown in Fig. \ref{fig:int_circuit}. 
As indicated by the insertion loss analysis and the PD responsivity, the estimated maximum photocurrent $I_{PD,max}$ is 110 $\mu A$.
Given a maximum targeted voltage of $V_{DD}=240\ mV$, the signal data rate of 5 GHz, and a target integration timestep $T$=60, we can derive the capacitor $C_{int}=I_{PD,max}T/(fV_{DD})=5500$ fF. 
Therefore, two foundry-compatible thin oxide capacitors with a capacitance range of 809 fF to 3.9 nF are connected to the PD's output.
Note that besides scaling up capacitors proportionally with $T$, one can equivalently consider scaling down laser power and thus $I_{PD,max}$ by a factor of $T$.
This can significantly reduce laser power but at the cost of a worse signal-to-noise ratio.
In our design, we maintain the same laser power and include the $T$ factor in the capacitance.

For a linear integrator response, multiple flipped capacitor pairs are connected in parallel to achieve a symmetric circuit topology.
To enable fast periodic reset, ten 40 nm n-channel and p-channel FETs are connected in parallel with the capacitor to ensure sufficient current driving capability for reset within a single baud time period.
This choice accounts for the possibility of both positive and negative source current flow from the balanced photodiode, ensuring effective reset regardless of signal polarity. 
For simplicity, only two of each type of FET are depicted in Fig. \ref{fig:int_circuit}.

Note that we prefer this capacitor-based design to an alternative operational amplifier (op-amp) based design due to efficiency considerations.
Integrators with an op-amp and a capacitive feedback loop show desired input/output impedance; however, they are more suitable for voltage integration tasks with notably increased chip space usage and power. 
In contrast, the capacitor-based design has near-zero power and is more suitable for our photocurrent accumulation mechanism in \ours.
\begin{figure}
    \centering
    \includegraphics[width=0.85\columnwidth]{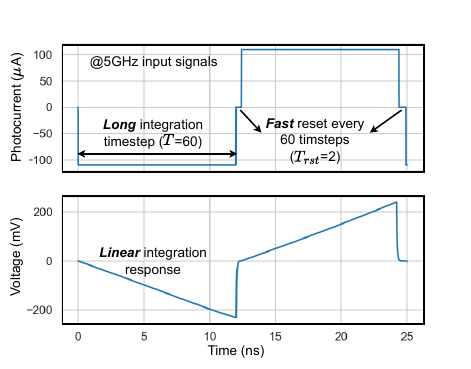}
    \caption{Simulated waveforms of the integrator unit.  
    Input photocurrent (\emph{Top}) and integrated voltage signal (\emph{Bottom}) show linear integration and rapid discharge (reset).
    }
    \label{fig:int_wav}
\end{figure}

\noindent\textbf{\underline{Integrator SPICE Simulation}} -- The integrator unit's simulation employs flipped capacitor pairs and 40 nm FETs, as previously mentioned.  
We simulated a maximum current of $\pm 110 \mu A$ over the entire integration period ($T$=60) to ensure saturation of the capacitor does not occur. 
The FET gates received 2.5 V for 120 ps, with additional rise and fall times of 40 ps, ensuring a complete reset within a timestep of $T_{rst}$=2. 
The waveforms for both the current signal and the integrated voltage signal are illustrated in Fig.~\ref{fig:int_wav}.  
Given the maximum anticipated current of $\pm$110 $\mu A$, we recorded peak voltages of approximately  $\mp$ 240 mV.

\noindent\textbf{\underline{Integrator Cost Analysis}} -- 
Our design shows a compact footprint of $A_{int}$=560 $\mu m^2$, a low power consumption of 0.3 mW, and a long integration timestep $T$=60, with a fast reset time $T_{rst}$ of 2 timesteps.
Note that the integrator arrays are shared across $C$ cores in a tile; the integrator area/power cost can be further amortized by a factor of $C$, leading to marginal hardware overhead at the system level.

\begin{figure}
    \centering 
    \includegraphics[width=0.85\columnwidth]{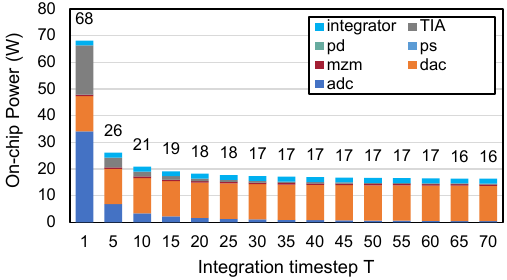}
    \vspace{-5pt}
    \caption{Impact of temporal integration timestep $T$ to the on-chip system power consumption for \ours-Custom-SL.
    Note that memory and off-chip laser are excluded.}
    \label{fig:IntegrationTimestep}
\end{figure}

\noindent\textbf{\underline{Integrator's Benefits to System Efficiency}} -- 
To justify the efficiency benefit by setting $T$ to 60, we simulate how timestep $T$ impacts the system power consumption when mapping a large matrix multiplication workload onto our architecture in Fig.~\ref{fig:IntegrationTimestep}.
The TIA/ADC sampling frequency can be scaled down proportionally by $T$ times, approximately leading to $T\times$ lower power. 
To keep ADC/TIA power less than 5\%, we set $T$ to 60 such that the on-chip power consumption can be drastically reduced from 68 W to 16 W, with the ADC/TIA bottleneck completely eliminated.

\section{Evaluation Results}
\label{sec:Results}
In this section, we will analyze the accuracy and hardware cost of our \ours architecture.
We focus on three variants of our \ours with different device configurations listed in Table~\ref{tab:device}.
\ours-Custom-SL is the fully-customized architecture settings used as our final design.
For a comprehensive evaluation of \ours-Custom-SL, we also incorporated the analysis of on-chip memory, considering its area and power impact ~\cite{NP_arXiv2023_Zhu}. 
Similar to ~\cite{NP_arXiv2023_Zhu}, the architecture has a 2MB global on-chip SRAM buffer and 4KB on-chip local SRAM buffer for each tile, designed to hold two $512\times512$ matrix multiplication workloads.
To summarize, \ours-Custom-SL consumes 321 mm$^2$ area, 17.5 W power at 5 GHz and $T$=60 integration timestep, and realizes 368.6 TOPS peak computing speed with 6-bit precision, 22.3 TOPS/W energy efficiency, and 1.2 TOPS/mm$^2$ compute density.

\begin{table*}[]
\centering
\small{
\caption{Component parameters used in three of \ours variants.
IL represents insertion loss. 
}
\label{tab:device}}
\resizebox{0.75\textwidth}{!}{
\begin{tabular}{|c|c|c|c|c|c|}
\hline
Device   & Parameter   & Value    & \begin{tabular}[c]{@{}c@{}}\ours \\Foundry \end{tabular}& \begin{tabular}[c]{@{}c@{}}\ours \\Foundry-SL \end{tabular} & \begin{tabular}[c]{@{}c@{}}\ours \\Custom-SL \end{tabular} \\ \hline
\multirow{3}{*}{DAC~\cite{NP_VLSI2020_Caragiulo}} & Precision     & 8-bit      & &   &        \\
                     & Power         & 50 mW(@14GSPS)  & \textcolor{red}{\large$\star$}& \textcolor{orange}{\large$\star$} & \textcolor{green}{\large$\star$}\\ 
                     & Area         & 11,000 $\mu m^2$  & & & \\\hline
\multirow{3}{*}{ADC~\cite{NP_ISSCC2022_Liu}} & Precision     & 8-bit     & &     &      \\
                     & Power         & 14.8 mW(@10GSPS)& \textcolor{red}{\large$\star$} & \textcolor{orange}{\large$\star$}& \textcolor{green}{\large$\star$}\\ 
                     & Area         & 2,850 $\mu m^2$ & && \\ 
                     \hline
\multirow{5}{*}{Foundry Photodetector~\cite{NP_OFC2018_Timurdogan,NP_OFC2020_Rakowski}}             & Power & 25 nW at -1 V& & &\\
                                          & Sensitivity & -27 dBm      & & &               \\
                                          & Area        & 16 $\times$20 $\mu m^2$  &\textcolor{red}{\large$\star$} & \textcolor{orange}{\large$\star$}  & \textcolor{green}{\large$\star$}                     \\
                                          & Bandwidth        & 27 GHz    & &   &                    \\
                                          & Responsivity        & 1.1 A/W   & &   &                     \\
                                          \hline
\multirow{3}{*}{TIA~\cite{NP_VLSI2018_Rakowski}} & Power        & 3 mW     & &   &               \\
                                          & Area        & $<$50 $\mu m^2$   &\textcolor{red}{\large$\star$} &   \textcolor{orange}{\large$\star$}  & \textcolor{green}{\large$\star$}              \\
                                          & Bandwidth        & 40 GHz   & &         &           \\
                                          \hline\hline
\multirow{6}{*}{Foundry MZM~\cite{NP_OFC2018_Timurdogan}}                      & Static power        & 70 nW & & & \\
                                          & IL        &  3 dB     & &  &               \\
                                          & Area        & 1600$\times$ 460 $\mu m^2$ & \textcolor{red}{\large$\star$}& &\\
                                          & EO bandwidth        & 12.5 GHz & & &\\
                                          & Modulation efficiency        & 450 fJ/bit & & &\\
                                          & Extinction ratio        & >15 dB& & &\\\hline
\multirow{6}{*}{\begin{tabular}[c]{@{}c@{}}Customized SL-MZM~\cite{NP_IEEEPTL2023_Anderson} \\(Fabricated at AIM)\end{tabular}}                      & Static power        & 70 nW at -3.5 V & & &\\
                                          & IL        &  6.4 dB     & &    &             \\
                                          & Area        & 250$\times$ 25 $\mu m^2$ & &\textcolor{orange}{\large$\star$} & \textcolor{green}{\large$\star$} \\
                                          & EO bandwidth        & 10 GHz (foreseeable)& & & \\
                                          & Modulation efficiency        & 50 fJ/bit& & & \\
                                          & Extinction ratio        & 6 dB& & & \\\hline \hline
\multirow{2}{*}{\begin{tabular}[c]{@{}c@{}}Foundry 2$\times$2 50:50 MMI~\cite{NP_OFC2020_Rakowski}\end{tabular}} & IL        & 0.11 dB     & \textcolor{red}{\large$\star$} & \textcolor{orange}{\large$\star$} &                 \\
                                          & Area        & 36$\times$10$\mu m^2$    & &    &               \\ \hline
\multirow{2}{*}{\begin{tabular}[c]{@{}c@{}}Customized 2$\times$2 50:50 \\Directional coupler \end{tabular}} & IL        & 0.05 dB     &  &  &  \textcolor{green}{\large$\star$}               \\
                                          & Area        & 31$\times$6.5$\mu m^2$    & &    &               \\ \hline\hline
\multirow{3}{*}{Foundry TO phase shifter~\cite{NP_OFC2020_Rakowski}}            & IL        & 0.03 dB      & &       &            \\
                                          & Area        & 75$\times$75 $\mu m^2$  & \textcolor{red}{\large$\star$} & \textcolor{orange}{\large$\star$}  &                 \\ 
                                           & Power        & $P_{\pi}$=7 mW   & &     &              \\ \hline
\multirow{3}{*}{Customized phase shifter}            & IL        & 0.05 dB    & &      &  \textcolor{green}{\large$\star$}             \\
                                          & Area        & 0.5$\times$33 $\mu m^2$  & &  &                  \\ 
                                           & Power        & $\sim0$ W  & &  &                  \\ \hline \hline                                          
\multirow{2}{*}{Customized 1$\times$10 splitter}                 & IL        & 0.199 dB       &  \textcolor{red}{\large$\star$}      &\textcolor{orange}{\large$\star$} &\textcolor{green}{\large$\star$}         \\
                                          & Area        & 34.6$\times$14.1$\mu m^2$   & & &                 \\ \hline
\multirow{2}{*}{Foundry 1$\times$2 50:50 MMI~\cite{NP_OFC2020_Rakowski}}                 & IL        &  0.1 dB          & \textcolor{red}{\large$\star$}&  \textcolor{orange}{\large$\star$} &  \textcolor{green}{\large$\star$}         \\
                                          & Area        & 22$\times$10 $\mu m^2$  & &    &               \\ \hline
\multirow{2}{*}{Foundry waveguide crossing~\cite{NP_OFC2020_Rakowski}}                 & IL        & 0.23dB       & \textcolor{red}{\large$\star$} & \textcolor{orange}{\large$\star$} & \textcolor{green}{\large$\star$}              \\
                                          & Area        & 8$\times$8 $\mu m^2$    & &  &               \\ \hline
Fiber/chip coupling & IL & 2 dB & \textcolor{red}{\large$\star$} & \textcolor{orange}{\large$\star$} &\textcolor{green}{\large$\star$} \\\hline
\multirow{1}{*}{Laser}                 
                                          & Wavelength        & 1550 nm  & \textcolor{red}{\large$\star$}&  \textcolor{orange}{\large$\star$} & \textcolor{green}{\large$\star$}                \\ \hline
\end{tabular}
}
\vspace{-10pt}
\end{table*}

\subsection{Accuracy Evaluation on Various Edge AI Workloads}
\begin{figure*}
    \centering
    \includegraphics[width=0.95\textwidth]{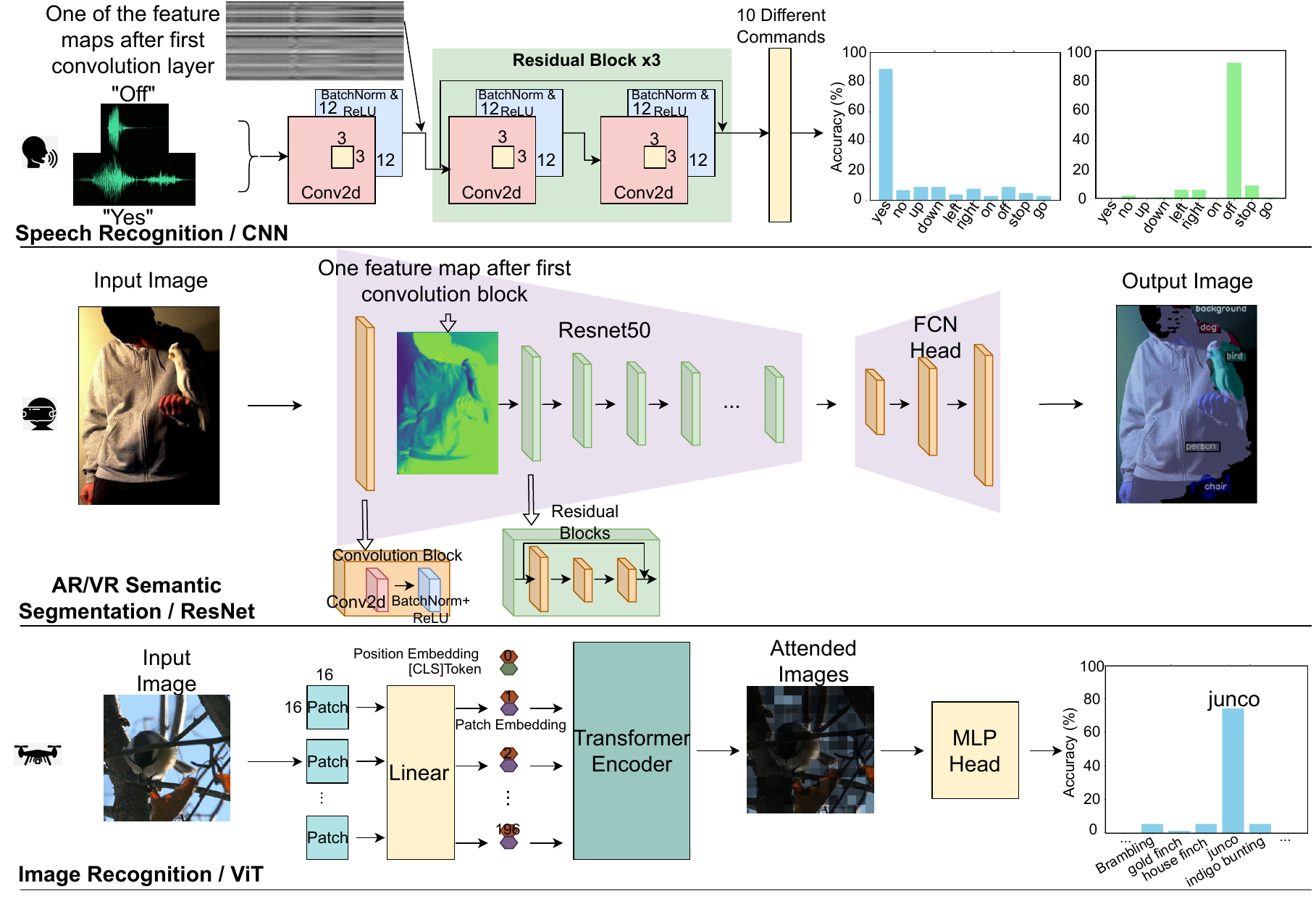}
    \caption{Evaluation of our \ours accelerator on three edge machine learning tasks, including image recognition, voice keyword spotting, and semantic segmentation on CNNs and Vision Transformers (ViT).
    All optical NNs are trained with 6-bit weight/activation quantization and hardware noise injections.}
    \label{fig:TaskDemo}
\end{figure*}

The performance of the proposed \ours is evaluated on real-world edge machine learning tasks, including a Vision Transformer (ViT) DeiT-Tiny~\cite{NN_ICML2021_Touvron} on image recognition on ImageNet-1k~\cite{NN_imagenet2009}, a convolutional neural network (CNN) on the AR/VR voice keyword spotting task on Google Speech Command dataset~\cite{NN_SpeechRecog2018}, and a FCN-ResNet50~\cite{NN_CVPR2015_Long} model on semantic segmentation on PASCAL VOC2012~\cite{NN_VOC2012}. 
Our evaluation covers both weight-static CNNs and Transformers with dynamic self-attention operations for both speech and vision tasks to demonstrate our versatility for diverse edge ML.
During model training, we adopt a hardware-aware training flow to consider the 6-bit weight/input quantization and hardware-measurement noises to guarantee a robust deployment on our photonic tensor cores.

\noindent\textbf{\underline{Noise/Quantization-Aware Training}} -- 
We adopt learnable step-size per-channel quantization~\cite{NN_ICLR2020_Esser} for both input operands $X$ and $Y$ and the output $S$.
For weight/activation quantization, the $i$-th channel of the quantized tensors is
\begin{equation}
    \label{eq:WeightQuantization}
    X_q^i=\calQ(X^i)= ([\texttt{clip(}X^i / \alpha^i + z^i, -2^{b-1}, 2^{b-1}-1)] - z^i)\cdot\alpha^i,
\end{equation}
where the scaling factor $\alpha^i$ and the zero point $z^i$ can be trained with gradient descent for the $i$-th channel/kernel.
The gradient of the non-differentiable rounding function can be estimated by using a straight-through estimator (STE).
After quantization, we also dynamically inject relative random Gaussian noises with a noise intensity of $\sigma$ to both input tensors in matrix multiplication, i.e., $\widetilde{X}_q=X_q + \Delta X$, where $\Delta X\sim \calN(0, (\sigma|X_q|)^2)$.

Figure~\ref{fig:TaskDemo} visualizes our proposed \ours on three representative edge AI workloads.
Table~\ref{tab:Accuracy} shows the task performance on each application with 6-bit weight/activation quantization and noise perturbations.
Our 6-bit quantized \ours can realize comparable recognition and segmentation performance on edge AI tasks.
\begin{table*}[]
\caption{Accuracy of \ours on 4 benchmarks with INT-6 weight/activation quantization and noise perturbation ($\sigma$=0.01).}
\label{tab:Accuracy}
\resizebox{0.82\textwidth}{!}{
\begin{tabular}{|c|c|c|c|c|}
\hline
Task                         & Dataset               & Model     & Fp32 Performance & INT6+Noise Acc \\ \hline
Image Recognition            & ImageNet-1k~\cite{NN_imagenet2009}         & DeiT-Tiny~\cite{NN_ICML2021_Touvron} & 0.722 (Accuracy)        & 0.712 (Accuracy)        \\ \hline
AR/VR Voice Keyword Spotting & Google Speech Command~\cite{NN_SpeechRecog2018} & CNN~\cite{NN_SpeechRecog2018}       & 0.957 (Accuracy)      & 0.929 (Accuracy)      \\ \hline
AR/VR Semantic Segmentation  & Pascal VOC2012~\cite{NN_VOC2012}        & FCN R-50-D8~\cite{NN_CVPR2015_Long}      & 52.28 (mIoU)         &  51.16 (mIoU)             \\ \hline
\end{tabular}
}
\end{table*}

\noindent\textbf{\underline{Noise Robustness Evaluation}} --
\begin{figure*}
    \centering
    \subfloat[]{
    \includegraphics[width=0.6\columnwidth]{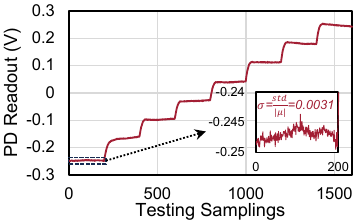}
    \label{fig:NoiseSweep}
    }
    \hspace{60pt}
    \subfloat[]{
    \includegraphics[width=0.8\columnwidth]{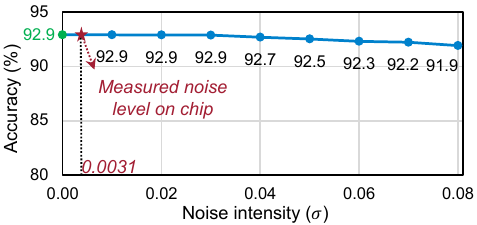}
    \label{fig:RobustnessAnalysis}
    }
    \caption{(a) Noise measurement in experimental chip testing of SL-MZM.
    (b) Inference accuracy evaluation on the CNN speech command benchmark with various noise intensities ($\sigma$) from 0 to 0.08.
    The model is trained with the noise-aware quantization method.
    The noise intensity (0.0031) observed in the SL-MZM chip testing shows negligible accuracy impact.}
    \label{fig:Rubustness}
\end{figure*}

To assess the robustness of our architecture against noise, we tested our speech recognition model with noise-aware training under various noise intensities injected in inference. Figure~\ref{fig:RobustnessAnalysis} indicates that our architecture demonstrates superior robustness to random noises. 
Even when increasing the relative noise intensity $\sigma$ from 0 to 0.08, the accuracy drops by only 1\%. 
Additionally, we measure the real noises in the chip testing in Fig.~\ref{fig:NoiseSweep}, which causes a negligible accuracy drop. 

\subsection{System Architecture-Level Performance Analysis}
As a case study, we configure our architecture with 6$\times$6 PTCs ($R=C=6$), and each PTC is of size $32\times 32$ ($K=32$), working at a clock rate of 5 GHz.
We give area and power estimation of our architecture.

\begin{figure}
    \centering
    \subfloat[]{\includegraphics[width=0.434\columnwidth]{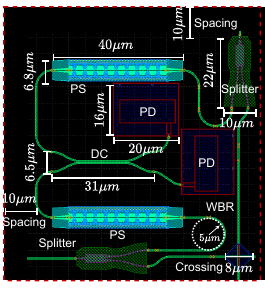}
    \label{fig:NodeLayoutR}
    }
    \subfloat[]{\includegraphics[width=0.532\columnwidth]{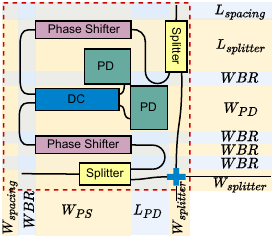}
    \label{fig:NodeLayoutB}
    }
    \vspace{-5pt}
    \caption{(a) Layout of one dot-product engine (node). 
    (b) Area breakdown for the node area $A_{node}$. 
    WBR is denoted as waveguide bending radius, we use 5 $\mu m$ as the WBR.
    }
    \label{fig:NodeLayout}
\end{figure}

\begin{figure}
    \centering
    \subfloat[]{\includegraphics[width=0.7\columnwidth]{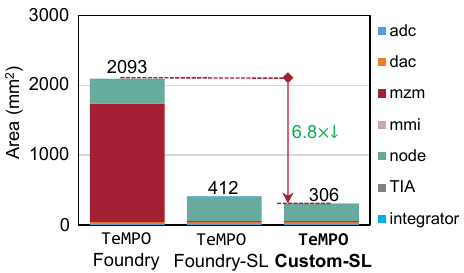}
    \label{fig:AreaBreakdown}
    }\\
    \vspace{-10pt}
    \subfloat[]{\includegraphics[width=0.7\columnwidth]{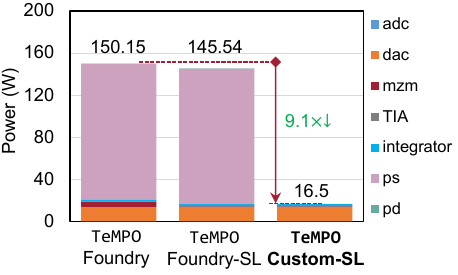}
    \label{fig:PowerBreakdown}
    }
    \vspace{-5pt}
    \caption{(a) Area and (b) on-chip power breakdown of our proposed \ours across 3 different device configurations (6$\times$6 PTCs, each with a size of 32$\times$32) working at 5 GHz and 1550 nm wavelength.
    Note that memory is excluded.
    \ours with customized devices achieves 6.8$\times$ smaller area and 9.1$\times$ lower power compared to Foundry PDKs.}
    \label{fig:AreaPowerBreakdown}
\end{figure}

\begin{figure}
    \centering
    \subfloat[]{\includegraphics[width=0.61\columnwidth]{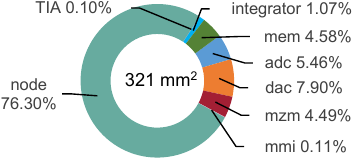}
    \label{fig:AreaPie}
    }\\
    \subfloat[]{\includegraphics[width=0.63\columnwidth]{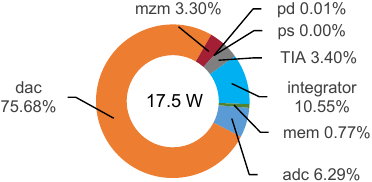}
    \label{fig:PowerPie}
    }
    \vspace{-5pt}
    \caption{(a) Area and (b) on-chip power breakdown of our \ours-custom-SL architecture (6$\times$6 PTCs, each with a size of 32$\times$32) working at 5 GHz and 1550 nm wavelength.
    Note that memory is included.
    }
    \label{fig:AreaPowerPie}
\end{figure}

\noindent\textbf{\underline{Area Cost}} --
The total area cost of a $K\times K$ PTC, including photonics and electronics, is estimated as follows
\begin{equation}
    \label{eq:AreaCost}
    \begin{aligned}
        A=&2K \cdot A_{DAC}+2K\cdot A_{MZM} + A_{1\times 2K~MMI}\\
    &+ K^2 (A_{node}+A_{int}+A_{TIA}+A_{ADC}),
    \end{aligned}
\end{equation}
where each node area in the crossbar can be estimated by the bounding box $A_{node}=(L_{splitter}+4\text{WBR}+W_{PD}+W_{splitter}+L_{spacing})(W_{splitter}+\text{WBR}+W_{PS}+L_{PD}+W_{spacing})$,
where WBR is the waveguide bending radius (set to 5 $\mu m$).
Figure~\ref{fig:NodeLayout} shows the details of how we derived the node area. 
We draw the layout in Fig.~\ref{fig:NodeLayoutR} and show the dimension calculation details in Fig.~\ref{fig:NodeLayoutB}.
Other area terms can be directly obtained from the device area specifications.
Note that the $1 \times 2K$ MMI is scaled based on our 1$\times$10 MMI design, assuming length/width is proportional to fanout.
Figure~\ref{fig:AreaBreakdown} shows the area comparison among 3 \ours variants.
With Foundry-based high-speed E-O MZM, the PTC area is bulky, where the MZMs took almost 81\% of the total circuit area.
With our compact slow-light MZMs, the total area is reduced by 6.8$\times$, while the MZMs only take 4.7\% of the total area.
Figure~\ref{fig:AreaPie} further includes on-chip memory in the breakdown.
Our customized architecture's area cost is 321 mm$^2$, where 76.3\% of the area is from the crossbar structure with minimum peripheral overhead from input encoding and data readout.

\noindent\textbf{\underline{Power Consumption}} --
We first give an analysis of the system-level on-chip power
\begin{equation}
\small
    \label{eq:Power}
    P=2K\cdot (P_{DAC}+\cdot P_{MZM})+K^2\cdot (2P_{PD}+P_{int}+P_{TIA}+P_{ADC}).
\end{equation}

The DAC power can be derived by $P_{DAC}=\frac{P_0b_0 2^bf}{2^{b_0}bf_s}$, where $P_0$ is the DAC power at $b_0$-bit precision and $f_s$ sampling rate, and $f$ is the clock frequency.
Other power terms can be directly obtained from the device power specification.

We emphasize the benefits of our multi-core architecture and temporal integration mechanism in power efficiency: 
\ding{202} Our multi-tile architecture can reduce the MZM and DAC power by a factor of $R$ for matrix $Y$ since the matrix $Y$ modulation components are shared across $R$ tiles before the on-chip waveguide broadcast, shown in Fig.~\ref{fig:ArchOverview}.
\ding{203} Multiple cores per tile share the same array of integrators, TIAs, and ADCs.
Meanwhile, as we analyzed in Section~\ref{sec:Integrator}, temporal integration can further reduce the TIA and ADC working frequency by a factor of $T$.
Hence, the power of TIA/ADC can be overall reduced by $CT$ times.

Figure~\ref{fig:PowerBreakdown} shows the power breakdown of the three variants of \ours.
Compared to the foundry MZM, which takes 450 fJ to encode each symbol, our designed SL-MZM only takes 50 fJ to encode each symbol, leading to an 89\% reduction in the input tensor modulation power consumption.
With time integration ($T=60$), the ADC/TIA power is reduced by 60$\times$, which becomes negligible (<5\%) in the system power.

\begin{figure*}[hbt!]
    \centering
    \subfloat[]{\includegraphics[width=0.21\textwidth]{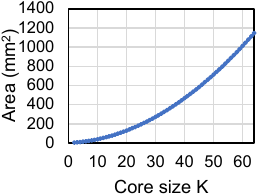}
    \label{fig:CoreArea}
    }
    \hspace{20pt}
    \subfloat[]{\includegraphics[width=0.2\textwidth]{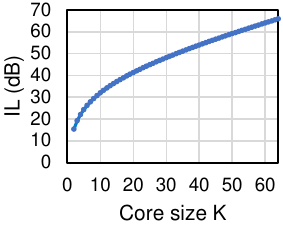}
    \label{fig:CoreIL}
    }\hspace{20pt}
    \subfloat[]{\includegraphics[width=0.205\textwidth]{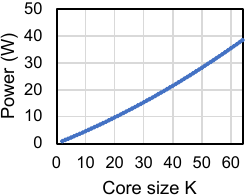}
    \label{fig:CorePower}
    }\\\vspace{-10pt}
    \subfloat[]{\includegraphics[width=0.212\textwidth]{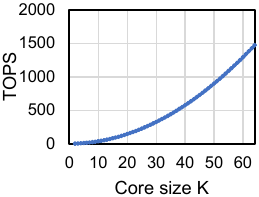}
    \label{fig:CoreTOPS}
    }\hspace{20pt}
    \subfloat[]{\includegraphics[width=0.203\textwidth]{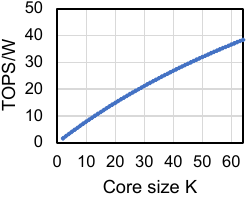}
    \label{fig:CoreTOPSW}
    }\hspace{20pt}
    \subfloat[]{\includegraphics[width=0.211\textwidth]{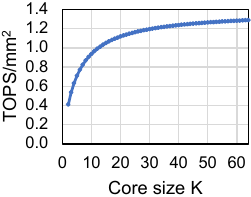}
    \label{fig:CoreTOPSmm2}
    }
    \caption{(a) Area, (b) IL, (c) power, (d) computing speed (TOPS), (e) energy efficiency (TOPS/W) and (f) compute density (TOPS/mm\textsuperscript{2}) with different PTC core size $K$ of our \ours-custom-SL (6$\times$6 cores) working at 5 GHz.
    }
    \label{fig:CoreScaling}
\end{figure*}

\begin{figure}[hbt!]
    \centering
    \includegraphics[width=\columnwidth]{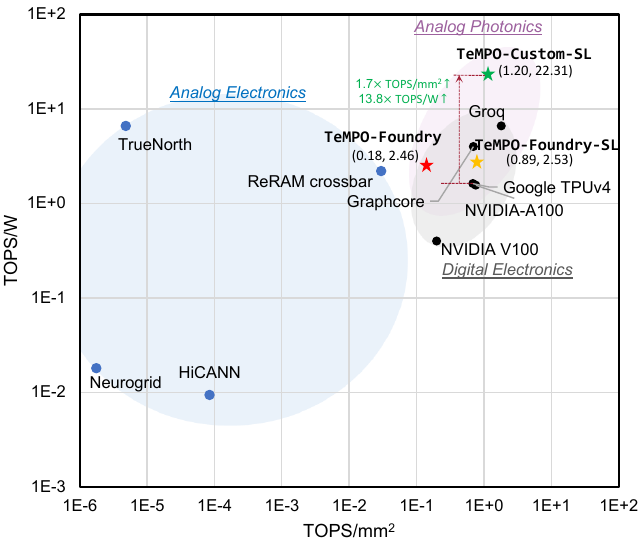}
    \caption{Compare digital electronics (NVIDIA V100~\cite{V100GPU}, A100 GPU~\cite{A100GPU}, Google TPUv4~\cite{TPUV4}, Groq~\cite{Groq_2020}, and Graphcore~\cite{Graphcore_2002}), analog electronics (IBM TrueNorth~\cite{TrueNorth_2019}, Neurogrid~\cite{Neurogrid_2014}, HiCANN~\cite{NN_ISCAS2010_Schemmel}, and ReRAM crossbar~\cite{HWA_VLSI2021_Giordano}), and our photonic AI hardware \ours in compute density (TOPS/mm\textsuperscript{2}) and energy efficiency (TOPS/W).
    Our \ours-Custom-SL with customized devices is at the Pareto front.}
    \label{fig:ParetoFront}
\end{figure}

Overall, our optimized \ours-Custom-SL architecture equipped with energy-efficient SL-MZMs, customized splitters, phase shifters, and temporal integrators can reduce the on-chip system-level power by 9.1$\times$ compared to foundry PDK variants.
Figure~\ref{fig:PowerPie} indicates \ours-Custom-SL consumes 17.5 W power while 76\% of power is from DACs.
As technology continues advancing, power-efficient DACs are expected to significantly boost the efficiency of \ours further.

\subsection{Tensor Core Efficiency and Scalability Analysis}
In this section, we show a thorough analysis of the scalability of one PTC with different core sizes $K$.
Besides area, insertion loss (IL), and power, we further define computing speed, energy efficiency, and compute density. 
To estimate the peak performance, we define the computing speed for each core as $2K^2fT/(T+T_{rst})$.
Note that the reset overhead is considered as a scaling factor $T/(T+T_{rst})$.
To evaluate the area efficiency, we adopt the metric of peak compute density, which measures how fast the hardware can compute per unit circuit area.
For a \ours architecture ($R\times C$ cores) with $K\times K$ PTC, the peak compute density is evaluated as $\frac{2K^2RCT}{fA(T+T_{rst})}$, where $f$ is the clock frequency (no higher than the maximum ADC sampling rate, i.e., $f\le f_{ADC,max}$).
The energy efficiency of the hardware is defined as $\frac{2K^2RC}{fP}$ if we ignore energy cost during reset as the accelerator is idle, which measures how much energy it consumes to finish one operation.

Our \ours architecture has $6\times6$ PTCs, and each PTC core size varies from $2\times2$ to $64\times64$. 
Figure~\ref{fig:CoreArea} shows a nearly quadratic area scaling since most of the area is attributed to the crossbar structure with quadratically many dot-product engines.
Figure~\ref{fig:CoreIL} shows almost linear insertion loss scaling as the number of crossings and splitters linearly increases with the core size $K$.
Hence, it is not efficient to use an overly large core size due to intractable insertion loss and laser power.
In Fig.~\ref{fig:CorePower}, we observe that power linearly scales with core size.
Since the hardware power is dominated by DAC and we have a linear number of DAC to encode input vectors.
Compared to quadratic power scaling in electronic circuits (as the transistor count quadratically increases with a larger $K$), this linear power scaling shows the advantage of photonic computing cores.
Figure~\ref{fig:CoreTOPS} shows the superior peak performance of our multi-core photonic accelerator.
With 5 GHz computing frequency and a core size of 30-40, \ours can potentially realize Peta operations per second (POPS)-level computing speed.
Thanks to the quadratically increasing computing speed and the linear power scaling, \ours shows a consistent efficiency boost with a larger core size in Fig.~\ref{fig:CoreTOPSW}.
In terms of compute density, we can obtain a higher density with a larger core size, as indicated by Fig.~\ref{fig:CoreTOPSmm2}.
We expect a higher compute density in the future with more compact coupler and photodetector designs as technology advances.
Overall, \ours shows good scalability to a larger core size.
The ultimate upper bound of core size is from the insertion loss, which can be largely relaxed with customized low-loss optical components.

\subsection{Efficiency Comparison with SoTA Accelerator Designs}
We compare our designs with state-of-the-art (SoTA) electronic digital computers, including GPU, TPU, ASIC, and analog neuromorphic processors, e.g., IBM TrueNorth.
We observe that our architecture \ours can realize competitive energy efficiency and compute density compared to state-of-the-art digital computers.
However, standard foundry PDK devices are not the most efficient designs for photonic computing.
By replacing the foundry MZM with our SL-MZM alone, we can boost the compute density from 0.18 (\ours-Foundry) to 0.89 (\ours-Foundry-SL) TOPS/mm$^2$.
With customized SL-MZM, splitters, and phase shifters, our fully customized \ours-Custom-SL pushes the Pareto frontier to a record high level. 
It achieves 22.3 TOPS/W and 1.2 TOPS/mm$^2$, outperforming the foundry PDK variant by 9.1$\times$ higher energy efficiency and 6.8$\times$ higher compute density, respectively. 
Compared to NVIDIA A100 GPU and Google TPUv4, \ours-Custom-SL shows 13.8$\times$ higher TOPS/W and 1.7$\times$ higher compute density, respectively.

\section{Conclusion}
\label{sec:Conclusion}
In this work, we present \ours, a time-multiplexed dynamic photonic tensor accelerator designed for energy-efficient edge AI applications. 
Through careful co-design across device, circuit, and architecture layers, \ours achieves significant performance improvements compared to state-of-the-art electronic accelerators. 
Key innovations include customized slow-light Mach-Zehnder modulator, optical splitter, and phase shifters for low-power dynamic tensor computation, analog domain accumulation via capacitive temporal integration to eliminate analog-to-digital conversion bottleneck, and a multi-core architecture for efficient hardware sharing. 
\ours demonstrates comparable task accuracy with 6-bit quantization to digital counterparts, superior noise tolerance, and a peak performance of 368.6 TOPS, energy efficiency of 22.3 TOPS/W, and compute density of 1.2 TOPS/mm$^2$, pushing the Pareto frontier for edge AI hardware. 
This work establishes a new frontier in energy-efficient analog AI hardware, paving the path for future electronic-photonic accelerators in ubiquitous edge AI applications.

\section{Data Availability}
The data that support the findings of this study are available
within the article.

\section*{References}
\vspace{-20pt}
%

\newpage
\end{document}